\documentclass[%
 reprint,
 amsmath,amssymb,
 aps,
prapplied,
longbibliography,
]{revtex4-2}

\usepackage[fleqn]{nccmath}
\usepackage{graphicx}
\usepackage{dcolumn}
\usepackage{bm}
\usepackage{xcolor}
\usepackage[normalem]{ulem}
\usepackage{hyperref}
\usepackage{footnote}
\usepackage[super]{nth}

\begin{document}

\preprint{APS/123-QED}

\title{Simultaneous Polarization Conversion and Anomalous Reflection \\ with Anisotropic Printed-Circuit-Board (PCB) Metagratings}

\author{Sharon Elad}
\altaffiliation{sharon.elad@campus.technion.ac.il}

\author{Ariel Epstein}
\email{epsteina@ee.technion.ac.il}

\affiliation{Andrew and Erna Viterbi Faculty of Electrical Engineering,\\
Technion - Israel Institute of Technology, Haifa 3200003, Israel.}

\date{\today}

\begin{abstract}
We present a semianalytical synthesis scheme for designing realistic printed-circuit-board- (PCB) based metagratings (MGs), capable of \emph{simultaneous} beam steering \emph{and} polarization conversion. These low-profile structures, comprised of sparse periodic arrangements of subwavelength polarizable particles (meta-atoms), oriented by insightful analytical models leading to fabrication-ready designs, have gained growing interest in recent years, demonstrating exceptional diffraction engineering capabilities with very high efficiencies. Herein, to facilitate cross-polarization coupling, we introduce a tilted dogbone meta-atom, going beyond the popular vertical (loaded wire) or recently proposed horizontal (dipole) configurations used in common PCB MG manifestations at microwaves, strictly susceptible to either transverse electric (TE) or transverse magnetic (TM) fields, respectively. As shown, the anisotropy introduced by the in-plane rotation of the dogbones, integrated into the MG analytical model in the form of a rotated dipole line, serves as an additional degree of freedom that can be leveraged to nonlocally manipulate the polarization and trajectory of the incident fields. Verified experimentally, the resultant semianalytical synthesis framework enables valuable physical insights, opening the door to integration of efficient anisotropic MGs in applications requiring both polarization and wavefront control, such as satellite, radar and advanced antenna systems.
\end{abstract}

\maketitle

\begin{figure*}
\centering
\includegraphics[width=2\columnwidth]{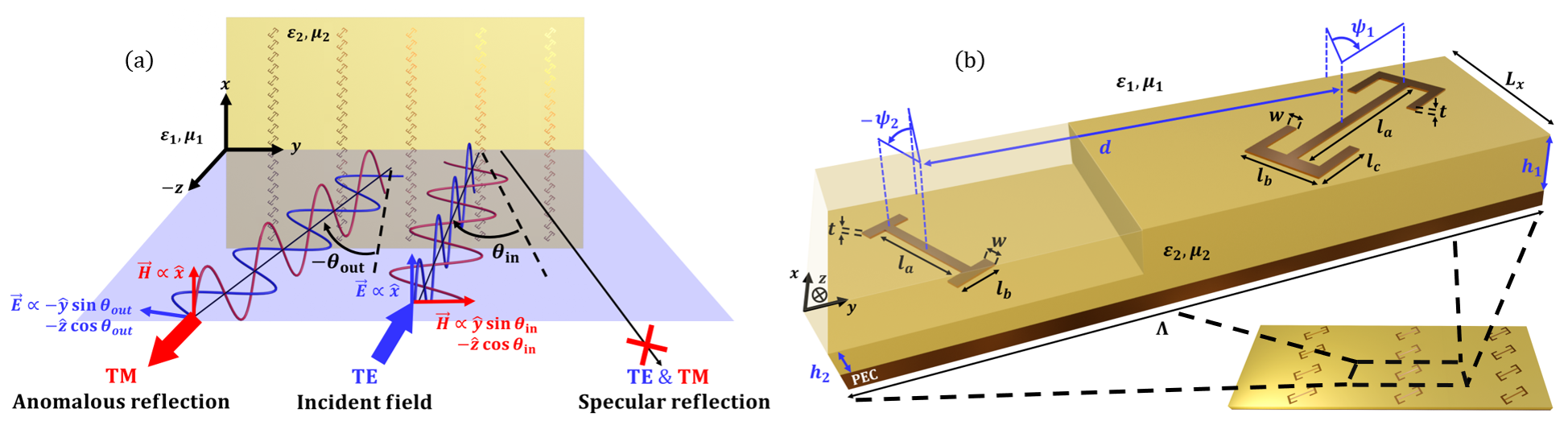}
\caption{Physical configuration of an anisotropic polarization- and wavefront-converting MG. (a) A TE-polarized plane wave impinging upon the MG from $\theta_\mathrm{in}$ is converted into a TM-polarized wave reflected towards $\theta_\mathrm{out}$, suppressing all other FB modes (both TE and TM). 
(b) Trimetric side view of a single ($L_x\!\times\!\Lambda$) unit cell, of the PCB-based MG (shown below). Implemented on a standard microwave-compatible substrate of width $h_1$, the considered PEC-backed ($z\!=\!0$) MG is surrounded by air ($z\!<\!-h_1$), and comprises two meta-atoms per unit cell; each features a thickness $t$, width $w$, primary length $l_a$, and secondary lengths (arms) $l_b$ and $l_c$ used to tune its polarizability (and subsequently the induced dipole moment). One dipole scatterer is situated on the dielectric-air interface at the center of the unit cell (at $(y,z)=(0,-h_1)$), and at a $d$ $y$-offset from it, the second one is ($h_1\!\geq\!h_2\!>\!0$) submerged within the substrate at a $h_2$ distance from the PEC (at $(y,z)=(d,-h_2)$); they are tilted by $\psi_1$ and $\psi_2$, respectively, with respect to the $x$ axis. Note (a) features only the dipole scatterers on the dialectic-air interface, as the others are (generally) embedded within the substrate.}
\label{Fig_config_and_general_UC}
\end{figure*}

\section{Introduction}
Metagratings (MGs) and metasurfaces (MSs) have manifested in recent years the herald of a new era in electromagnetics and photonics, succeeding the decades-long research on periodic structures and artificial materials. They, unlike their bulky  metamaterial (MM) predecessors, are thin metallo-dielectric structures composed of subwavelength-size polarizable particles termed meta-atoms (MAs). Deliberate tailoring of their degrees of freedom (DOFs) yields unique wave phenomena, otherwise unachievable with conventional material slabs \cite{GlybovskiMetasurfacesVisible, Raadi2022MetagratingsManipulation}.

Albeit proven to be very useful in the past two decades \cite{Iyer2020MetamaterialsDirections}, MS synthesis process typically necessitates substantial time-consuming full-wave simulations, to transform the abstract (homogenized) generalized sheet transition conditions (GSTCs) based macroscopic constituents \cite{Tretyakov2003AnalyticalElectromagnetics, Kuester2003AveragedMetafilm} into actual (physically realizable) densely packed (microscopic) MA arrangements \cite{Epstein2016HuygensApplications}. MGs, on the other hand, which feature sparsely distributed MAs, cannot be thus homogenized. Instead, to design them, the mutual coupling between their polarizable elements is incorporated via an extensive analytical model, forming a synthesis scheme that directly involves their structural DOFs \cite{Radi2017MetagratingsControl}. This detailed semianalytical approach oftentimes yields fabrication-ready design specifications, circumventing the need for any further optimizations, while offering unique analytical access to the underlying physics, stemming from the insightful interplay between their (relatively few) available DOFs and the desired functionality \cite{Epstein2017UnveilingAnalysis, Rabinovich2018AnalyticalReflection}.

In recent years, numerous demonstrations of highly efficient MGs performing a broad range of diffraction engineering functionalities, in various operating frequency bands, have been reported \cite{Radi2017MetagratingsControl,Memarian2017WidebandangleBS, Sell2017Large-AngleGeometries, Wong2018PerfectMetasurface, Yang2018FreeformSteering, Dong2020EfficientMetagratings, Raadi2022MetagratingsManipulation}. Specifically at microwave frequencies, we have developed a semianalytical synthesis scheme for implementing MGs using printed-circuit-board (PCB) compatible configurations, relying on capacitively-loaded conducting strips as MAs \cite{Epstein2017UnveilingAnalysis, Rabinovich2018AnalyticalReflection}. Suitable analytical formulation, modelling the excited elongated strips as canonical line sources radiating in stratified media (the PCB stack), enables resolution of the MA coordinates and load impedance to generate specific interference patterns upon excitation, thereby meeting prescribed functionality requirements \cite{Rabinovich2019ArbitraryMetagratings}. Subsequent model extensions of these PCB MGs led to the development of various advanced wavefront manipulating functionalities in reflection \cite{Rabinovich2018AnalyticalReflection, Popov2018ControllingMetagratings, Popov2019ConstructingFreedom, Rabinovich2019ExperimentalMetagratings, Xu2021DualWires} and transmission \cite{Rabinovich2019PrintedExperiment, Rabinovich2019ArbitraryMetagratings, Xu2021AnalysisApproach}, effectively tackling a variety of challenges in dynamic beam steering \cite{Casolaro2019DynamicMetagratings, Popov2021NonManipulations, Xu2022WideAntenna}, waveguides \cite{Biniashvili2021Eliminating, Killamsetty2021MetagratingsExperiment}, antenna applications \cite{Popov2020ConformalManipulation, Xu2022ExtremeOptimization, Kerzhner2022MetagratingApplications, Hu2023LargeSwitching, Hu2023CavityPatterns}, and radar-cross-section (RCS) reduction \cite{Yashno2022LargeAbsorption, Yashno2023BroadDiffusers, Tan2023DesignExperiment} to name a few.

However, these MG configurations can manipulate in a controlled manner only transverse electric (TE) polarized fields (where the incident electric field is parallel to the loaded wire MAs). Although there are a few demonstrations of PCB-based MGs receptive to the orthogonal transverse magnetic (TM) polarization \cite{Memarian2017WidebandangleBS, Ra'di2018reconfigurable}, the utilized MA geometries (narrow slots, essentially) inherently
restrict their operation to a single polarization as well. Very recently, a PCB-based MG device capable of a dual-polarized response has been proposed, extending the notion of the loaded wire configuration by introducing independent orthogonal dogbone elements to interact with TM polarized fields (in addition to the TE-susceptible loaded wires) \cite{Shklarsh2024SemianalyticallyMetagratings}. Another recent report uses a different set of orthogonal meta-atoms in conjunction with a polarization selective mirror for similar functionalities \cite{Nguyen2024DesignReflection}, cross-polarization coupling, though, is not facilitated by these geometries, still limiting their use. As an alternative architecture, all metallic MG devices suitable for operating in either single \cite{Rahmanzadeh2020PerfectMetagrating, Rajabalipanah2021AnalyticalMetagratings} or dual \cite{Rabinovich2020Dual-PolarizedReflection} polarization modes have also been proposed, however for applications requiring light-weight and small volume devices, PCB MGs usually have the upper hand.

Whether all-metallic or PCB-based, these MGs feature only negligible interaction between the two polarizations, and therefore cannot perform effective polarization manipulation. This hinders their integration in applications such as satellite communication \cite{Wang2021NovelMetasurfaces}, multiplexing in advanced antenna systems \cite{Guo2016AdvancesApplications}, and RCS reduction \cite{Ameri2019UltraMetasurfaces}, where polarization conversion capabilities are required and mere operation in either polarization is insufficient. Indeed, MSs performing such simultaneous polarization and trajectory manipulation have been previously proposed \cite{Asadchy2016PerfectMetasurfaces, Yepes2021PerfectSlab}. However, due to the need to obey the homogenization approximation underlying MS design, they require design and realization of many different densely packed anisotropic MAs, which may prove to be a relatively challenging task. 
Clearly, devising PCB MGs that would house polarization manipulation capabilities alongside anomalous beam deflection would form an attractive alternative to yield simple and effective solutions to relevant applications.

In this paper, we bridge this gap by breaking the convention of the popular loaded wire MA configuration governing PCB MG architecture to date, realizing a simple yet highly efficient anisotropic MG device capable of simultaneous beam steering and polarization conversion. To this end, we consider a MA geometry in which the polarizable elements are allowed to be rotated in plane along arbitrary angles \cite{Elad2023AnisotropicReflection} (in contrast to previous work, where they were restricted to vertical \cite{Rabinovich2018AnalyticalReflection} or horizontal \cite{Shklarsh2024SemianalyticallyMetagratings} orientations). When excited, such a configuration manifests a MG composed of periodically arranged lines of rotated dipoles. The corresponding DOFs, i.e., MA rotation angles, coordinates, and induced currents, are incorporated into the semianalytical synthesis scheme following the universal PCB MG design approach, establishing a generalized (polarization inclusive) design procedure. The developed theoretical framework, applied to realize a PCB MG which converts a TE-polarized incident wave into a TM-polarized one while anomalously reflecting it, is verified through both full-wave simulation and experiment. Importantly, closed-form analytical expressions dictating the values of the various MG parameters are derived, leading to insightful physical interpretation of the resultant MG operation and underlying mechanisms. These introduced anisotropic architecture and accompanying semianalytical extended model lay the cornerstone for developing advanced polarized beam manipulating PCB MG devices, further broadening the applicability range of these promising complex media apparatuses.

\section{Theory} \label{sec_Theory}

\subsection{Formulation} \label{subsec_Formulation}

We consider a MG configuration with two subwavelength-size, electrically polarizable, passive MAs per $L_x\!\times\!\Lambda$ unit-cell \footnote{While the model can be formulated for an arbitrary number of MAs as in \cite{Rabinovich2019ArbitraryMetagratings}, we focus here, as a fundamental case study, on anisotropic MGs with precisely two MAs per period. As shall be shown in Section \ref{subsec_PolConv_&_AnoRef}, these provide the minimal number of DOFs required for realizing the desired simultaneous polarization conversion and beam deflection.}, where $L_x$ and $\Lambda$ denote the grating periodicities along $x$ and $y$, respectively [Fig. \ref{Fig_config_and_general_UC}(b)]. Time harmonic field dependency $e^{j\omega t}$ is assumed and suppressed, defining the operating frequency $f=\omega /2\pi$ and wavelength $\lambda\!=\!c_0/f$, where $c_0$ is the speed of light in vacuum. The free space and substrate regions are indicated as per $i=1$ and $i=2$, respectively, and correspondingly, we denote the wave impedances $\eta_i\!=\!\sqrt{\mu_i/\varepsilon_i}$ and wavenumbers $k_i\!=\!2\pi f\sqrt{\mu_i\varepsilon_i}$ in each region.

Within each period, two (metallic) elements are positioned at $z\!=\!-h_1$ and $z\!=\!-h_2$ ($0\!<\!h_2\!\leq\!h_1$) below a perfect electric conductor (PEC) ground plane lying at $z\!=\!0$. They are implemented on a dielectric substrate with permittivity $\varepsilon_2\!=\!\varepsilon_\mathrm{sub}$, surrounded by a $\varepsilon_1\!=\!\varepsilon_\mathrm{air}$ medium in the entire $z\!<\!-h_1$ half-space [Fig. \ref{Fig_config_and_general_UC}]. The first element (at $z\!=\!-h_1$) is centered at $y=0$ with respect to the unit cell, and the other one is at $y=d$, where $|d|<\Lambda/2$. They are rotated in-plane by $\psi_1$ and $\psi_2$, respectively, with values ranging from $-\!\pi/2$ to $\pi/2$, and feature conductor width $w$ and thickness $t$. As shall be shown in Section \ref{subsec_Analytical_insights}, it is this rotation that provides the physical mechanism for controlling the cross-polarization coupling.

Given a plane wave excitation with an arbitrary polarization, impinging from $z\!=\!-\infty$ upon the structure at a $\theta_\mathrm{in}$ incidence angle with respect to the normal [Fig. \ref{Fig_config_and_general_UC}(a)], we strive to tie the MG five structural DOFs ($h_1, h_2, \psi_1, \psi_2$, and $d$) to a user prescribed functionality, i.e. a specific simultaneous polarization conversion and anomalous reflection scenario. To this end, we derive a general analytical model solving the scattering problem for a given such anisotropic MG, to be utilized, in turn, for the inverse design of a PCB-compatible composite performing the required polarized beamforming.
In this regard, and to avoid possible ambiguities, we note that TE polarization refers herein to fields that are polarized along the $x$ axis ($\boldsymbol{E}\cdot\boldsymbol{\hat{y}}=\boldsymbol{E}\cdot\boldsymbol{\hat{z}}=\boldsymbol{H}\cdot\boldsymbol{\hat{x}}=0$), while TM polarization refers to the orthogonal set, in which the electric field vector resides in the $\widehat{yz}$ plane ($\boldsymbol{H}\cdot\boldsymbol{\hat{y}}=\boldsymbol{H}\cdot\boldsymbol{\hat{z}}=\boldsymbol{E}\cdot\boldsymbol{\hat{x}}=0$).

In accordance with the superposition principle, the total fields $\boldsymbol{E}^\mathrm{tot}$ may be calculated as a sum of two contributions: the primary external fields $\boldsymbol{E}^\mathrm{ext}$ evaluated in the absence of the MG scatterers, i.e. including only the illuminating field and its reflection off the PEC-backed homogeneous dielectric substrate; and the secondary fields $\boldsymbol{E}^\mathrm{MG}$ generated by the perturbating currents these external fields induce within the MG polarizable particles \cite{Epstein2017UnveilingAnalysis}.

Correspondingly, for a given incoming plane wave with amplitudes $E^{\mathrm{TE}}_{\mathrm{in}}$ and $E^{\mathrm{TM}}_{\mathrm{in}}$ for the TE and TM components, arriving from $z\!=\!-\infty$ at a $\theta_\mathrm{in}$ incidence angle with respect to the normal of the device, i.e.
\begin{equation} \label{eq:E_in}
\begin{split}
   & \boldsymbol{E}^\mathrm{in}\left(y,z\right) \\
   & \!=\!\left[\hat{x}E_{\mathrm{in}}^{\mathrm{TE}}\!-\!\left(\hat{y}\cos\theta_{0,1} \!-\!\hat{z}\sin\theta_{0,1} \right)E_{\mathrm{in}}^{\mathrm{TM}}\right] e^{-jk_{t_{0}}y}e^{-j\beta_{0,1}z},
\end{split}
\end{equation}
the external fields below the device ($z\!<\!-h_1$) read
\begin{equation} \label{eq:E^ext_1}
\begin{split}
   & \boldsymbol{E}_{1}^{\mathrm{ext}}\left(y,z\right) \\
   & =\!\hat{x}E_{\mathrm{in}}^{\mathrm{TE}}\left(e^{-j\beta_{0,1}z}\!+\!R_{0}^{\mathrm{TE}}e^{j\beta_{0,1}\left(z\!+\!2h_{1}\right)}\right)e^{-jk_{t_{0}}y} \\
   & -\!\hat{y}\cos\theta_{0,1}E_{\mathrm{in}}^{\mathrm{TM}}\left(e^{-j\beta_{0,1}z}\!-\!R_{0}^{\mathrm{TM}}e^{j\beta_{0,1}\left(z\!+\!2h_{1}\right)}\right)e^{-jk_{t_{0}}y} \\
   & +\!\hat{z}\sin\theta_{0,1}E_{\mathrm{in}}^{\mathrm{TM}}\left(e^{-j\beta_{0,1}z}\!+\!R_{0}^{\mathrm{TM}}e^{j\beta_{0,1}\left(z\!+\!2h_{1}\right)}\right)e^{-jk_{t_{0}}y},
\end{split}
\end{equation}
and in the region between the PEC and the air-substrate interface ($-h_1\!<\!z\!<\!0$, within the dielectric)
\begin{equation} \label{eq:E^ext_2}
\begin{split}
   & \boldsymbol{E}_{2}^{\mathrm{ext}}\left(y,z\right) \\
   & -\!\hat{x}E_{\mathrm{in}}^{\mathrm{TE}}T_{0}^{\mathrm{TE}}\frac{\sin\left(\beta_{0,2}z\right)}{\sin\left(\beta_{0,2}h_{1}\right)}e^{j\beta_{0,1}h_{1}}e^{-jk_{t_{0}}y} \\ 
   & +\!j\hat{y}\cos\theta_{0,2}\frac{\eta_{2}}{\eta_{1}}E_{\mathrm{in}}^{\mathrm{TM}}T_{0}^{\mathrm{TM}}\frac{\cos\left(\beta_{0,2}z\right)}{\sin\left(\beta_{0,2}h_{1}\right)}e^{j\beta_{0,1}h_{1}}e^{-jk_{t_{0}}y} \\ 
   & -\!\hat{z}\sin\theta_{0,2}\frac{\eta_{2}}{\eta_{1}}E_{\mathrm{in}}^{\mathrm{TM}}T_{0}^{\mathrm{TM}}\frac{\sin\left(\beta_{0,2}z\right)}{\sin\left(\beta_{0,2}h_{1}\right)}e^{j\beta_{0,1}h_{1}}e^{-jk_{t_{0}}y}.
\end{split}
\end{equation}
where $R_0^\mathrm{P}$ and $T_0^\mathrm{P}$ are, respectively, the (TE and TM) reflection and transmission coefficients given by \cite{Rabinovich2018AnalyticalReflection,Pozar2011MicrowaveEngineering}
\begin{equation} \label{eq:R0T0}
\begin{split}
   & R_{0}^{\mathrm{P}}=\frac{j\gamma_{0}^{\mathrm{P}}\tan\left(\beta_{0,2}h_{1}\right)-1}{j\gamma_{0}^{\mathrm{P}}\tan\left(\beta_{0,2}h_{1}\right)+1} \\
   & T_{0}^{\mathrm{P}}=1+R_{0}^{\mathrm{P}}=\frac{2j\gamma_{0}^{\mathrm{P}}\tan\left(\beta_{0,2}h_{1}\right)}{j\gamma_{0}^{\mathrm{P}}\tan\left(\beta_{0,2}h_{1}\right)+1},
\end{split}
\end{equation}
necessitated due to the presence of the air/dielectric and PEC ground interfaces, which evoke an inherent impedance mismatch leading to multiple reflections. For readability, we use the superscript $\mathrm{P}$ to denote polarization affiliation, where $\mathrm{P}$ may be substituted by $\mathrm{TE}$ or $\mathrm{TM}$ to obtain the results corresponding to either polarization.

Accordingly, the TE and TM wave impedance ratios are, respectively, $\gamma_{0}^{\mathrm{TE}}\!=\!Z_{0,2}^{\mathrm{TE}}/Z_{0,1}^{\mathrm{TE}}$ and $\gamma_{0}^{\mathrm{TM}}\!=\!Z_{0,1}^{\mathrm{TM}}/Z_{0,2}^{\mathrm{TM}}$, where
\begin{equation} \label{eq:Z0}
\begin{split}
   & Z_{0,i}^{\mathrm{TE}}\!=\!\eta_{i}k_i/\beta_{0,i}, \ 
    Z_{0,i}^{\mathrm{TM}}\!=\!\eta_{i}\beta_{0,i}/k_i,
\end{split}
\end{equation}
with $i\!\in\!\{1,2\}$ indicating either of the two media. As in \cite{Rabinovich2018AnalyticalReflection}, we define $\theta_{0,1}\!=\!\theta_\mathrm{in}$ as the incidence angle, and $\theta_{0,2}$ is related to it via Snell's law $k_1 \sin\theta_{0,1} \!=\! k_2 \sin\theta_{0,2}$. Correspondingly, the transverse and longitudinal wavenumbers are defined, respectively, as per
\begin{equation} \label{eq:kt0_b0}
\begin{split}
   & k_{t_0}\!=\!k_i \sin\theta_{0,i}, \ \beta_{0,i}\!=\!\sqrt{k_{i}^{2}-k_{t_{0}}^{2}}\!=\!k_i\cos\theta_{0,i}.
\end{split}
\end{equation}

As aforementioned, subject to the external fields Eqs. \eqref{eq:E^ext_1}-\eqref{eq:E^ext_2}, dipole moments $I\ell$ would be induced within the MG tilted passive polarizable MAs, aligned with their primary axes. Thus, the current density of a single and isolated $L_x$-periodic line of such $\psi$-rotated dipoles centered around the origin $(y',z')\!=\!(0,0)$ -- hereinafter referred to as a rotated dipole line (RDL) -- reads
\begin{equation} \label{eq:J^RDL}
\begin{split}
    & \boldsymbol{J}^{\mathrm{RDL}}(x,y,z)\! \\
    & \ \ =\boldsymbol{J}^{\mathrm{LS}}\left(\boldsymbol{r}\right)\cos\psi+\boldsymbol{J}^{\mathrm{DL}}\left(\boldsymbol{r}\right)\sin\psi \\
    & \ \ =\!\left(\hat{x}\cos\psi\!+\!\hat{y}\sin\psi\right)I\ell \delta(y)\delta(z)\sum_{n\in\mathbb{Z}}\delta\left(x\!-\!n L_{x}\right).
\end{split}
\end{equation}
Due to the introduced anisotropy, Eq. \eqref{eq:J^RDL} features two ($x$ and $y$) vector components, attributed, respectively, to the LS (line-source, $\psi\!=\!0$) \cite{Epstein2017UnveilingAnalysis} and DL (dipole-line, $\psi\!=\!\pi/2$) \cite{Shklarsh2024SemianalyticallyMetagratings} sub-constituents \cite{Felsen1973RadiationWaves}, out of which this RDL entity is composed of. We conform to the standard MG design approach, setting $L_{x}\!=\!\lambda/4\!\ll\!\lambda$, effectively constituting a 2D ($\partial/\partial x \approx 0$) configuration \cite{Epstein2017UnveilingAnalysis}. These perturbating currents [Eq. \eqref{eq:J^RDL}] further induce a secondary field (see Appendix \ref{Appndx_A}), reading
\begin{equation} \label{eq:E^RDL}
\begin{split}
    & \boldsymbol{E}^{\mathrm{RDL}}\left(y\!-\!y',z\!-\!z';\tilde{I},\psi\right) \\
    & =-\hat{x}\frac{k\eta}{4}\tilde{I}\cos\psi H_{0}^{(2)}\left(k\left|\boldsymbol{r}\!-\!\boldsymbol{r}'\right|\right) \\
    & \ \ \ +\frac{\eta}{4k}\tilde{I}\sin\psi\left[\hat{y}\frac{\partial^{2}}{\partial z^{2}}\!-\!\hat{z}\frac{\partial^{2}}{\partial y\partial z}\right]H_{0}^{(2)}\left(k\left|\boldsymbol{r}\!-\!\boldsymbol{r}'\right|\right).
\end{split}
\end{equation}
Herein, $\tilde{I}\!\overset{\triangle}{=}\!I\ell/L_{x}$ denotes the dipole moment per unit length, and $H_0^{(2)}(\Omega)$ is the $0$th order Hankel function of the second kind given as a function of $\left|\boldsymbol{r}\!-\!\boldsymbol{r}'\right|\!=\!\sqrt{(y\!-\!y')^{2}\!+\!(z\!-\!z')^{2}}$. Serving as the RDL Green's function in homogeneous medium, Eq. \eqref{eq:E^RDL} indicates \emph{both} a TE ($E_z=0$, first term) \emph{and} a TM ($H_z=0$, second term) response, where scattering into \emph{both} polarization components occurs \emph{only} for nontrivial $\psi\notin\{0,\pm\pi/2\}$.

Our configuration is thusly composed of two $\Lambda$-periodic RDL-gratings (along $y$); the first is situated at the air-dielectric interface $z\!=\!-h_1$, and the second is embedded within the substrate on the (virtual) interface $z\!=\!-h_2\!\leq\!-h_1$ [Fig. \ref{Fig_config_and_general_UC}(b)]. Each one of these grids independently induces a secondary set of fields associated with Eq.  \eqref{eq:E^RDL}, which, due to the $y$-periodicity and the Floquet-Bloch (FB) theorem, may be expressed as (two independent sets of) an infinite number of FB-harmonics \cite{Rabinovich2019ArbitraryMetagratings}. By decomposing the fields into these FB modal components, considering forward and backward waves multiply reflected between the configuration interfaces, and imposing the electromagnetic configuration constraints - the PEC boundary conditions at $z=0$, continuity conditions at the various interfaces, and source conditions around the polarizable RDLs (acting as secondary sources) - one arrives at the following expressions for the total (modal) secondary fields (Appendix \ref{Appndx_B}). In the three disjoint regions: $1$, below the device ($z\!<\!-h_{1}$); $2^-$, between the two arrays ($-h_{1}\!<\!z\!<\!-h_2$); and $2^+$, between the PEC and the adjacent MA-array ($-h_{2}\!<\!z\!<\!0$); these fields read, respectively   
\begin{widetext}
    \begin{equation} \label{eq:E^MG_1}
    \begin{split}
       & E_{m,1}^{\mathrm{MG}}\left(y,z\right) =\hat{x}\frac{1}{\Lambda}\left[\tilde{I}_{1}\cos\psi_{1}B_{m,1}^{\mathrm{TE},\left(1\right)}+\tilde{I}_{2}\cos\psi_{2}\hat{T}_{m}^{\mathrm{TE}}B_{m,2^{-}}^{\mathrm{TE},\left(2\right)}e^{jk_{t_{m}}d}\right]e^{j\beta_{m,1}z}e^{-jk_{t_{m}}y} \\
       & \ \ \ \ \ \ \ \ \ \ \ \ \ \ \ +\hat{y}\frac{1}{\Lambda}\left[\tilde{I}_{1}\sin\psi_{1}B_{m,1}^{\mathrm{TM},\left(1\right)}+\tilde{I}_{2}\sin\psi_{2}\hat{T}_{m}^{\mathrm{TM}}B_{m,2^{-}}^{\mathrm{TM},\left(2\right)}e^{jk_{t_{m}}d}\right]e^{j\beta_{m,1}z}e^{-jk_{t_{m}}y} \\
       & \ \ \ \ \ \ \ \ \ \ \ \ \ \ \ +\hat{z}\frac{1}{\Lambda}\left[\tilde{I}_{1}\sin\psi_{1}B_{m,1}^{\mathrm{TM},\left(1\right)}+\tilde{I}_{2}\sin\psi_{2}\hat{T}_{m}^{\mathrm{TM}}B_{m,2^{-}}^{\mathrm{TM},\left(2\right)}e^{jk_{t_{m}}d}\right]\tan\theta_{m,1}e^{j\beta_{m,1}z}e^{-jk_{t_{m}}y}; \\
       &
    \end{split}
    \end{equation}
    
    \begin{equation} \label{eq:E^MG_2}
    \begin{split}
       & E_{m,2^-}^{\mathrm{MG}}\left(y,z\right) =-\hat{x}\frac{1}{\Lambda}\left[\tilde{I}_{1}\cos\psi_{1}A_{m,2^{-}}^{\mathrm{TE},\left(1\right)}\sin\left(\beta_{m,2}z\right)\right. \\
       & \ \ \ \ \ \ \ \ \ \ \ \ \ \ \ \ \ \ \ \ \ \ \ \ \ \ \ \left.-\tilde{I}_{2}\cos\psi_{2}B_{m,2^{-}}^{\mathrm{TE},\left(2\right)}e^{jk_{t_{m}}d}e^{-j\beta_{m,1}h_{1}}\left(e^{j\beta_{m,2}\left(z+h_{1}\right)}\!+\!\hat{R}_{m}^{\mathrm{TE}}e^{-j\beta_{m,2}\left(z+h_{1}\right)}\right)\right]e^{-jk_{t_{m}}y} \\
       & \ \ \ \ \ \ \ \ \ \ \ \ \ \ \ \ \ \ \ -\hat{y}\frac{1}{\Lambda}\left[\tilde{I}_{1}\sin\psi_{1}A_{m,2^{-}}^{\mathrm{TM},\left(1\right)}\sin\left(\beta_{m,2}z\right)\right. \\
       & \ \ \ \ \ \ \ \ \ \ \ \ \ \ \ \ \ \ \ \ \ \ \ \ \ \ \ \left.-\tilde{I}_{2}\sin\psi_{2}B_{m,2^{-}}^{\mathrm{TM},\left(2\right)}e^{jk_{t_{m}}d}e^{-j\beta_{m,1}h_{1}}\left(e^{j\beta_{m,2}\left(z+h_{1}\right)}\!+\!\hat{R}_{m}^{\mathrm{TM}}e^{-j\beta_{m,2}\left(z+h_{1}\right)}\right)\right]e^{-jk_{t_{m}}y} \\
       & \ \ \ \ \ \ \ \ \ \ \ \ \ \ \ \ \ \ \ +\hat{z}\frac{j}{\Lambda}\left[\tilde{I}_{1}\sin\psi_{1}A_{m,2^{-}}^{\mathrm{TM},\left(1\right)}\cos\left(\beta_{m,2}z\right)\right. \\
       & \ \ \ \ \ \ \ \ \ \ \ \ \ \ \ \ \ \ \ \ \ \ \ \ \ \ \ \left.-j\tilde{I}_{2}\sin\psi_{2}B_{m,2^{-}}^{\mathrm{TM},\left(2\right)}e^{jk_{t_{m}}d}e^{-j\beta_{m,1}h_{1}}\left(e^{j\beta_{m,2}\left(z+h_{1}\right)}\!-\!\hat{R}_{m}^{\mathrm{TM}}e^{-j\beta_{m,2}\left(z+h_{1}\right)}\right)\right]\tan\theta_{m,2}e^{-jk_{t_{m}}y}; \\
       &
    \end{split}
    \end{equation}
    
    \begin{equation} \label{eq:E^MG_3}
    \begin{split}
       & E_{m,2^+}^{\mathrm{MG}}\left(y,z\right) =-\hat{x}\frac{1}{\Lambda}\left[\tilde{I}_{1}\cos\psi_{1}A_{m,2^{-}}^{\mathrm{TE},\left(1\right)}+\tilde{I}_{2}\cos\psi_{2}A_{m,2^{+}}^{\mathrm{TE},\left(2\right)}e^{jk_{t_{m}}d}e^{-j\beta_{m,2}h_{2}}\right]\sin\left(\beta_{m,2}z\right)e^{-jk_{t_{m}}y} \\
       & \ \ \ \ \ \ \ \ \ \ \ \ \ \ \ \ \ \ \ -\hat{y}\frac{1}{\Lambda}\left[\tilde{I}_{1}\sin\psi_{1}A_{m,2^{-}}^{\mathrm{TM},\left(1\right)}+\tilde{I}_{2}\sin\psi_{2}A_{m,2^{+}}^{\mathrm{TM},\left(2\right)}e^{jk_{t_{m}}d}e^{-j\beta_{m,2}h_{2}}\right]\sin\left(\beta_{m,2}z\right)e^{-jk_{t_{m}}y} \\
       & \ \ \ \ \ \ \ \ \ \ \ \ \ \ \ \ \ \ \ +\hat{z}\frac{j}{\Lambda}\left[\tilde{I}_{1}\sin\psi_{1}A_{m,2^{-}}^{\mathrm{TM},\left(1\right)}+\tilde{I}_{2}\sin\psi_{2}A_{m,2^{+}}^{\mathrm{TM},\left(2\right)}e^{jk_{t_{m}}d}e^{-j\beta_{m,2}h_{2}}\right]\tan\theta_{m,2}\cos\left(\beta_{m,2}z\right)e^{-jk_{t_{m}}y},
    \end{split}
    \end{equation}
\end{widetext}

\begin{figure}\label{Fig_ab_coeffs}
\centering
\includegraphics[width=\linewidth]{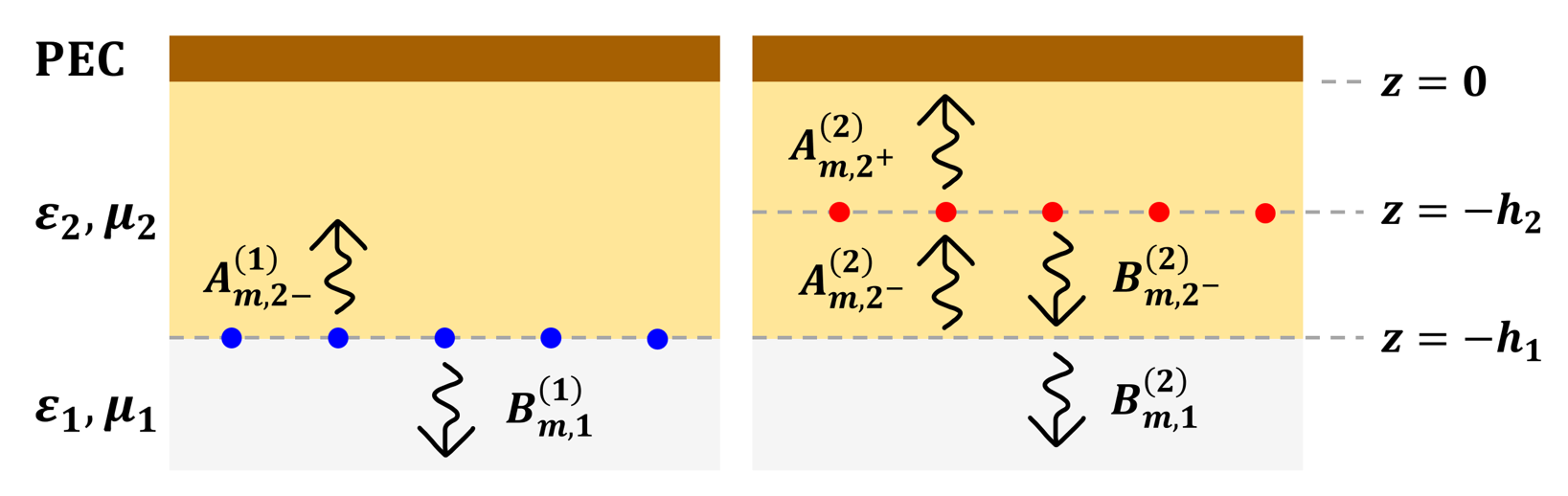}
\caption{Schematic front view of the MG featured in Fig. \ref{Fig_config_and_general_UC}. Adopting a stratified media modal-analysis approach to account for all multiple reflections, the TE and TM grid-induced fields of each dipole scatterer array (source) are represented by the forward and backward wave amplitude coefficients, determined by the boundary conditions. Whereas merely two coefficients per polarization are required in the case of source $(1)$, located on the dielectric-air interface (a), the immersed source $(2)$ requires double as much coefficients, due to reflections in the middle virtual layer (b).}
\end{figure}

In consistency with the notations featured in \cite{Rabinovich2019ArbitraryMetagratings} for the TE polarized case, the amplitudes $A^{\mathrm{P},(k)}_{m,i}$ and $B^{\mathrm{P},(k)}_{m,i}$ (featuring impedance units) of the modal field components in Eqs. \eqref{eq:E^MG_1}-\eqref{eq:E^MG_3}, correspond to the (TE and TM) forward and backward waves, respectively, within each of the three regions ($i\in\{1,2^-,2^+\}$), generated by each of the two RDL-grating sources ($k\in\{1,2\}$) [Fig. \ref{Fig_ab_coeffs}]. 
The explicit expressions for these amplitudes and relevant derivations are provided in Appendix \ref{Appndx_B}, wherein it is shown that the inter-relation between the various amplitudes can be used to reduce the ones in Eqs. \eqref{eq:E^MG_1}-\eqref{eq:E^MG_3}. Likewise, the utilized modal backward local reflection and transmission coefficients at $z=-h_1$ above, is defined as per
\begin{equation} \label{eq:RT_bw}
\begin{split}
    & \hat{R}_{m,2^{-}}^{\mathrm{P}}\!=\!A_{m,2^{-}}^{\mathrm{P},(2)}/B_{m,2^{-}}^{\mathrm{P},(2)}\!=\!\frac{Z_{m,1}^{\mathrm{P}}-Z_{m,2}^{\mathrm{P}}}{Z_{m,1}^{\mathrm{P}}+Z_{m,2}^{\mathrm{P}}} \\
    & \hat{T}_{m,2^{-}}^{\mathrm{P}}\!=\!B_{m,1}^{\mathrm{P},(2)}/B_{m,2^{-}}^{\mathrm{P},(2)}\!=\!1+\hat{R}_{m}^{\mathrm{P}},
\end{split}
\end{equation}
These are given in terms of the wave impedances of the $m^\mathrm{th}$ mode in the $i^\mathrm{th}$ medium, defined as per
\begin{equation} \label{eq:Zm}
\begin{split}
    & Z_{m,i}^{\mathrm{TE}}\!=\!\eta_{i}k_i/\beta_{m,i}, \ Z_{m,i}^{\mathrm{TM}}\!=\!\eta_{i}\beta_{m,i}/k_i,
\end{split}
\end{equation}
constituting a generalization of Eq. \eqref{eq:Z0}. The corresponding transverse and longitudinal wavenumbers are defined as per
\begin{equation} \label{eq:ktm&betami}
\begin{split}
   & k_{t_{m}}\!\overset{\triangle}{=}\!k_1\left(m\lambda/\Lambda+\sin\theta_{\mathrm{in}}\right)\!\overset{\triangle}{=}\!k_{i}\sin\theta_{m,i} \\
   & \beta_{m,i}\!=\!\sqrt{k_{i}^{2}-k_{t_{m}}^{2}}\overset{\triangle}{=}k_{i}\cos\theta_{m,i}, \ \mathfrak{Im}\left\{ \beta_{m,i}\right\} \!<\!0,
\end{split}
\end{equation}
where each $\theta_{m,i}$ pair is again related via Snell's law, altogether manifesting $k_{t_m}\!\overset{\triangle}{=}k_{t_{m,1}}\!=\!k_{t_{m,2}}$ (consistent with the definition Eq. \eqref{eq:kt0_b0} for $m=0$).

Finally, for a given incident field [Eq. \eqref{eq:E_in}] impinging upon a given anisotropic MG configuration (with DOFs $d,h_1,h_2,\psi_1,\psi_2,\varepsilon_1,\varepsilon_2$) as in Fig. \ref{Fig_config_and_general_UC}(a), and the correspondingly (yet to be evaluated) induced dipole moments per unit length $\tilde{I}_{1}$ and $\tilde{I}_{2}$, the total scattered fields (at $z\!<\!0$) are given via superposition as a sum of the external fields Eqs. \eqref{eq:E^ext_1}-\eqref{eq:E^ext_2} and the contributions from all the grating induced FB-modes Eqs. 
 \eqref{eq:E^MG_3}-\eqref{eq:E^MG_1}, namely
\begin{equation} \label{eq:E^tot}
\begin{split}
   & \boldsymbol{E}^{\mathrm{tot}}(y,z)=\boldsymbol{E}^{\mathrm{ext}}(y,z)+\sum_{m\in\mathbb{Z}}\boldsymbol{E}_{m}^{\mathrm{MG}}(y,z).
\end{split}
\end{equation}
As will be shown in Section \ref{subsec_PolConv_&_AnoRef}, these fields can be explicitly expressed as functions of the DOFs, to establish the accommodation of the minimal number of propagating FB modes necessary for beam steering and polarization conversion functionalities. Furthermore, in the limiting case $\varepsilon_1 \rightarrow \varepsilon_2$, these external and grid induced fields precisely reduce to those reported in \cite{Elad2023AnisotropicReflection} for a conceptual free-standing MG (no dielectric), as expected.

\subsection{Simultaneous polarization conversion \\ and anomalous reflection} \label{subsec_PolConv_&_AnoRef}

Now that we have formulated explicit terms for the total fields scattered off the MG [Eq. \eqref{eq:E^tot}], expressed via the the MG geometrical DOFs and the (yet to be determined) dipole moments per unit length, for a given (arbitrarily polarized) plane wave excitation, we proceed to establish analytical constraints tying these parameters to a user prescribed functionality involving simultaneous beam steering and polarization conversion. As a fundamental case study, we consider the simplest scenario for such polarized anomalous reflection MG. Specifically, we restrict ourselves to $\theta_\mathrm{in}$ and $\theta_\mathrm{out}$ values such that only two FB modes ($m=0$, corresponding to the spurious specular reflection, and $m=-1$, associated with the desired anomalous one), both TE and TM, will be propagating, and all else will be evanescent \cite{Radi2017MetagratingsControl}, \cite{Rabinovich2018AnalyticalReflection}, setting the period correspondingly as $\lambda/\Lambda\!=\!|\sin\left(\theta_{\mathrm{in}}\right)\!-\!\sin\left(\theta_{\mathrm{out}}\right)|$ as per the FB theorem. 

To attain perfect anomalous reflection, i.e. redirect all the incident power from $\theta_\mathrm{in}$ towards the anomalous $\theta_\mathrm{out}$ angle, we first wish to prevent any power from being coupled to the fundamental ($0^\mathrm{th}$ order) TE or TM modes in Eq. \eqref{eq:E^tot} \textcolor{black}{\footnote{\textcolor{black}{Note that due to the anisotropic nature of the rotated MAs, the incident field generally induces dipoles with non-vanishing projections on both the $x$ and $y$ axes, namely, generating scattered FB modes of both TE and TM polarizations. This is true even if the incident wave is strictly single-polarized (i.e., even if $E_\mathrm{in}^\mathrm{TE}=0$ or $E_\mathrm{in}^\mathrm{TM}=0$), as shall be demonstrated in Section \ref{sec_Results_and_Discussion}.}}}. This is done by demanding destructive interference between the incident wave's specular reflection off the PEC-backed dielectric and the $0^\mathrm{th}$ order FB-mode generated by the MG, which are associated, respectively, with the right-hand terms in Eqs. \eqref{eq:E^ext_1} and \eqref{eq:E^MG_1} for $m\!=\!0$. Altogether, formulating this destructive interference requirement for the two polarizations independently provides two equations requiring a specific relation between the induced dipole moments per unit length $\tilde{I}_{1}$ and $\tilde{I}_{2}$ to the MA locations and orientation DOFs ($h_1, h_2, \psi_1, \psi_2$ and $d$). These relations are given by
\begin{equation} \label{eq:I1I2}
\begin{split}
   & \left(\!\begin{array}{c}
\tilde{I}_{1}\\
\tilde{I}_{2}
\end{array}\!\right)\!=\Lambda\mathbf{Z}^{-1}\left(\!\begin{array}{c}
E_{\mathrm{in}}^{\mathrm{TE}}R_{0}^{\mathrm{TE}}\\
E_{\mathrm{in}}^{\mathrm{TM}}R_{0}^{\mathrm{TM}}\cos\theta_{\mathrm{in}}
\end{array}\!\right)e^{2j\beta_{0,1}h_1},
\end{split}
\end{equation}
constituting our \nth{1} and \nth{2} synthesis constraints, with the impedance matrix $\mathbf{Z}_{2\times2}$ defined as per (Appendix \ref{Appndx_B})
\begin{equation} \label{eq:M}
\begin{split}
    & \mathbf{Z}=\left(\begin{array}{cc}
B_{0,1}^{\mathrm{TE},(1)}\cos\psi_{1} & B_{0,1}^{\mathrm{TE},(2)}\cos\psi_{2}e^{jk_{t_{0}}d}\\
B_{0,1}^{\mathrm{TM},(1)}\sin\psi_{1} & B_{0,1}^{\mathrm{TM},(2)}\sin\psi_{2}e^{jk_{t_{0}}d}
\end{array}\right).
\end{split}
\end{equation}
The first and second rows of $\mathbf{Z}$ represent the TE and TM response of the MG, respectively, while the first and second columns are associated with the first ($h_1,\psi_1$) and second ($h_2,\psi_2,d$) MA contribution.

Subsequently, we quantify the MG polarization conversion efficiency by considering the TE and TM reflected fractional power
{\begin{equation} \label{eq:eta_-1}
\begin{split}
    & \eta_{-1}^{\mathrm{TE}} \overset{\triangle}{=} P_{-1}^{\mathrm{TE}}/(P_{-1}^{\mathrm{TE}}+P_{-1}^{\mathrm{TM}}) \\ & \eta_{-1}^{\mathrm{TM}} \overset{\triangle}{=} P_{-1}^{\mathrm{TM}}/(P_{-1}^{\mathrm{TE}}+P_{-1}^{\mathrm{TM}}),
\end{split}
\end{equation}}
where, following Eq. \eqref{eq:E^MG_1}, $P_{-1}^{\mathrm{P}}$ are the TE and TM power carried by the $m=-1$ mode, defined as
{\begin{equation} \label{eq:pTEpTM}
\begin{split}
    & P_{-1}^{\mathrm{TE}}\!=\!\frac{\left|\tilde{I}_{1}\cos\psi_{1}B_{-1,1}^{\mathrm{TE},\left(1\right)}+\tilde{I}_{2}\sin\psi_{2}B_{-1,1}^{\mathrm{TE},\left(2\right)}e^{jk_{t_{-1}}d}\right|^{2}}{2\Lambda Z_{-1,1}^{\mathrm{TE}}}, \\
    & P_{-1}^{\mathrm{TM}}\!=\!\frac{\left|\tilde{I}_{1}\sin\psi_{1}B_{-1,1}^{\mathrm{TM},\left(1\right)}+\tilde{I}_{2}\sin\psi_{2}B_{-1,1}^{\mathrm{TM},\left(2\right)}e^{jk_{t_{-1}}d}\right|^{2}}{2\Lambda Z_{-1,1}^{\mathrm{TM}}}.
\end{split}
\end{equation}}
These definitions [Eq. \eqref{eq:eta_-1}] are utilized to set the output polarization ratio as needed (i.e., as prescribed by the user), yielding our \nth{3} constraint.

Next, as we are striving to realize a passive and lossless MG configuration, mere suppression of the spurious modes and power balancing the two propagating ones, though key, is in itself insufficient, since, in principle, some of the power may be absorbed or amplified in the process. Thus, we introduce a \nth{4} constraint, stipulating that the total reflected real power must be precisely equal to the incident power. In other words, we require that all the incident power (neither less nor more) would indeed be channeled to the desired anomalous $m\!=\!-\!1$ reflection mode with the desirable TE or TM polarization. This can be formulated as
\begin{equation} \label{eq:GlobPowerCons}
\begin{split}
    & P_{-1}^{\mathrm{TE}}+P_{-1}^{\mathrm{TM}}\!=\!\frac{\Lambda}{2\eta_{1}}\cos\theta_{\mathrm{in}}\left(\left|E_{\mathrm{in}}^{\mathrm{TE}}\right|^{2}+\left|E_{\mathrm{in}}^{\mathrm{TM}}\right|^{2}\right).
\end{split}
\end{equation}

Nevertheless, this global power conservation condition Eq. \eqref{eq:GlobPowerCons} does not consider the possible interplay (power exchange) between the two MA arrays. These might feature balanced gain and loss imperceptible to Eq. \eqref{eq:GlobPowerCons}, which would then require (undesirably complex) active or resistive MAs for their physical realization. Hence, to ensure that \emph{both} scatterers are passive and purely reactive, we embrace the approach featured in \cite{Rabinovich2019PrintedExperiment}; we derive an explicit term for the net real power crossing some plane $-h_1\!<\!z\!=\!z_p\!<\!-h_2$ between these two elements $P_{z}^{\mathrm{tot}}(z_{p})$, integrating the $z$ component of Poynting’s vector over a single period $|y|\!<\!\Lambda/2$ and require that this spatially independent term will vanish,
\begin{equation}\label{eq:P_z^tot(mid)}
\begin{aligned}
   & P_{z}^{\mathrm{tot}}(z_{p})\overset{\triangle}{=}\frac{1}{2} \Re\left\{ \hat{z} \cdot  \intop_{-\Lambda/2}^{\Lambda/2}\!d y \ \boldsymbol{E}^{\mathrm{tot}}\!\times\!\boldsymbol{H}^{\mathrm{tot}^{*}} \! \right\}\!=\!0,
\end{aligned}
\end{equation}
thus manifesting our \nth{5} constraint.

Lastly, striving to perform the specified deflection and polarization conversion functionality at the highest efficacy, we further impose a \nth{6} constraint, aiming to reduce Ohmic losses within the copper traces as much as possible. These will emerge due to the finite resistivity presented by realistic copper, the effects of which were neglected so far, but should be accounted for if optimal designs are sought \cite{Rabinovich2019ArbitraryMetagratings}. Considering that the loss is proportional to the induced currents within the MA elements \cite{Epstein2017UnveilingAnalysis}, this conductor loss minimization requirement can be formulated as
\begin{equation} \label{eq:P_loss}
\begin{split}
    & P_\mathrm{loss}\overset{\triangle}{=}\frac{\eta_{1}}{L_{x}\Lambda}\left(\left|\tilde{I}_{1}\right|^{2}+\left|\tilde{I}_{2}\right|^{2}\right)\rightarrow \ 0,
\end{split}
\end{equation}

All in all, for a given incident field with amplitudes $E^\mathrm{TE}_\mathrm{in}$ and $E^\mathrm{TM}_\mathrm{in}$ for the two polarization components, and angle of incidence $\theta_\mathrm{in}$; and user prescribed TE or TM output polarization ratios $\eta_{-1}^\mathrm{TE}$, $\eta_{-1}^\mathrm{TM}$ and desired anomalous reflection angle $\theta_\mathrm{out}$, we utilize the \emph{general} formalism showcased herein, which includes 7 available DOFs ($\tilde{I}_1,\tilde{I}_2,h_1,h_2,\psi_1,\psi_2,d$), and establish 6 fully analytical synthesis constraints. The \nth{1} and \nth{2} constraints include cancellation of the two polarization components of the specular reflection using Eq. \eqref{eq:I1I2}; the \nth{3} constraint sets the power ratio between the two polarization components of the anomalous reflection (the only remaining propagating mode) Eq. \eqref{eq:eta_-1}; the \nth{4} is the global power conservation condition Eq. \eqref{eq:GlobPowerCons}; the \nth{5} enforces power balancing between the two MAs (per unit-cell) via Eq. \eqref{eq:P_z^tot(mid)}, guaranteeing a possible realization with passive and lossless elements; and finally, via our \nth{6} constraint Eq. \eqref{eq:P_loss}, low loss and high efficiency is obtained by virtue of minimizing the conductor losses.

Indeed, it is now clear that a MG featuring a single MA per period would not possess sufficient number of DOFs to meet all the required constraints for the functionality in mind. This is the reason for our choice to include two different MAs in each period, providing 7 DOFs to address the 6 stipulated constraints. As will be demonstrated in the next section, including two MAs per period as proposed [Fig. \ref{Fig_config_and_general_UC}], makes the MG well compatible for the designated polarized beam deflection functionality, constituting the simplest MG configuration (minimal number of MAs per period) in this regard.

With all these constraints formulated as a set of six explicit nonlinear equations \eqref{eq:I1I2}, \eqref{eq:GlobPowerCons}, \eqref{eq:P_z^tot(mid)}-\eqref{eq:P_loss}, the various MG design parameters may be resolved via a standard mathematical library function \footnote{As in \cite{Rabinovich2019ArbitraryMetagratings}, appropriate DOF values for thus realizing a prescribed scattering functionality are found by running the MATLAB library function $\mathtt{lsqnonlin}$ ca. 50 times with random initial values.}, and thereupon identify the combination that yields the highest anomalous reflection efficiency with the prescribed polarization ratio. The final synthesis stage requires finding of actual geometries (physical realization) for the MAs that would enable the evolution of the prescribed currents in response to the incident field; this will be addressed in Section \ref{subsec_Experiment}.

\section{Results and Discussion} \label{sec_Results_and_Discussion}

\subsection{Perfect TE-to-TM polarization conversion \\ and anomalous reflection} \label{subsec_Perfect_TE-to-TM}

To verify the general theoretical scheme developed in Section \ref{subsec_PolConv_&_AnoRef}, we apply it to design a PCB-based MG, aimed at reflecting an illuminating TE-polarized plane wave impinging upon it at a $\theta_\mathrm{in}\!=\!10^\circ$ incidence angle, towards an anomalous $\theta_\mathrm{out}\!=\!-60^\circ$ trajectory, whilst \emph{simultaneously} converting its polarization to TM ($E^{\mathrm{TE}}_{\mathrm{out}}\!=\!0$) at the output. \textcolor{black}{As laid out in Section \ref{sec_Theory}, the anisotropic rotated MAs incorporated to introduce coupling between the TE and TM field polarizations may give rise to both TE and TM propagating FB modes in all the relevant (allowed according to the FB theorem) set of angles – even if the incident field only features TE-polarized components. In particular, in the case considered herein, these propagating modes correspond to two (i.e., TE and TM) waves in the specular reflection direction ($m=0$ mode), and two (i.e., TE and TM) waves in the anomalous reflection trajectory ($m=-1$). Correspondingly, the synthesis goal would be to suppress the three spurious modes and direct all of the incident TE power towards the anomalous TM mode. Specifically, this implies utilization of the devised model with $\eta_{-1}^\mathrm{TE}=0$ in Eq. \eqref{eq:eta_-1}}.

To this end, we choose the low-loss microwave laminate Rogers RO4350B as a substrate, compatible for the selected operating frequency of $20 \ \mathrm{GHz}$ ($\lambda\!=\!14.989 \ \mathrm{mm}$), featuring permittivity of $\varepsilon_2\!=\!\varepsilon_\mathrm{sub}\!=\!3.66\varepsilon_0$. First, we find the 7 MG optimal DOF values for this specific case study; these include the induced dipole moments per unit length $\tilde{I}_1$ and $\tilde{I}_2$ on the two MAs, the respective horizontal offset $d$, the tilt angles $\psi_1$ and $\psi_2$, and the distances from the PEC $h_1$ and $h_2$ [Fig. \ref{Fig_config_and_general_UC}]. By virtue of our 6 synthesis constraints formulated as a set of explicit nonlinear equations, these parameters may be readily resolved via code as done in \cite{Rabinovich2019ArbitraryMetagratings}. Therein, appropriate DOFs' values for a given scattering problem were found by running the MATLAB library function $\mathtt{lsqnonlin}$ ca. 50 times with random initial values, and thereupon the combination yielding the highest anomalous reflection efficiency with the prescribed polarization ratio was identified and used for the design.

\begin{figure} 
\centering
\includegraphics[width=\linewidth]{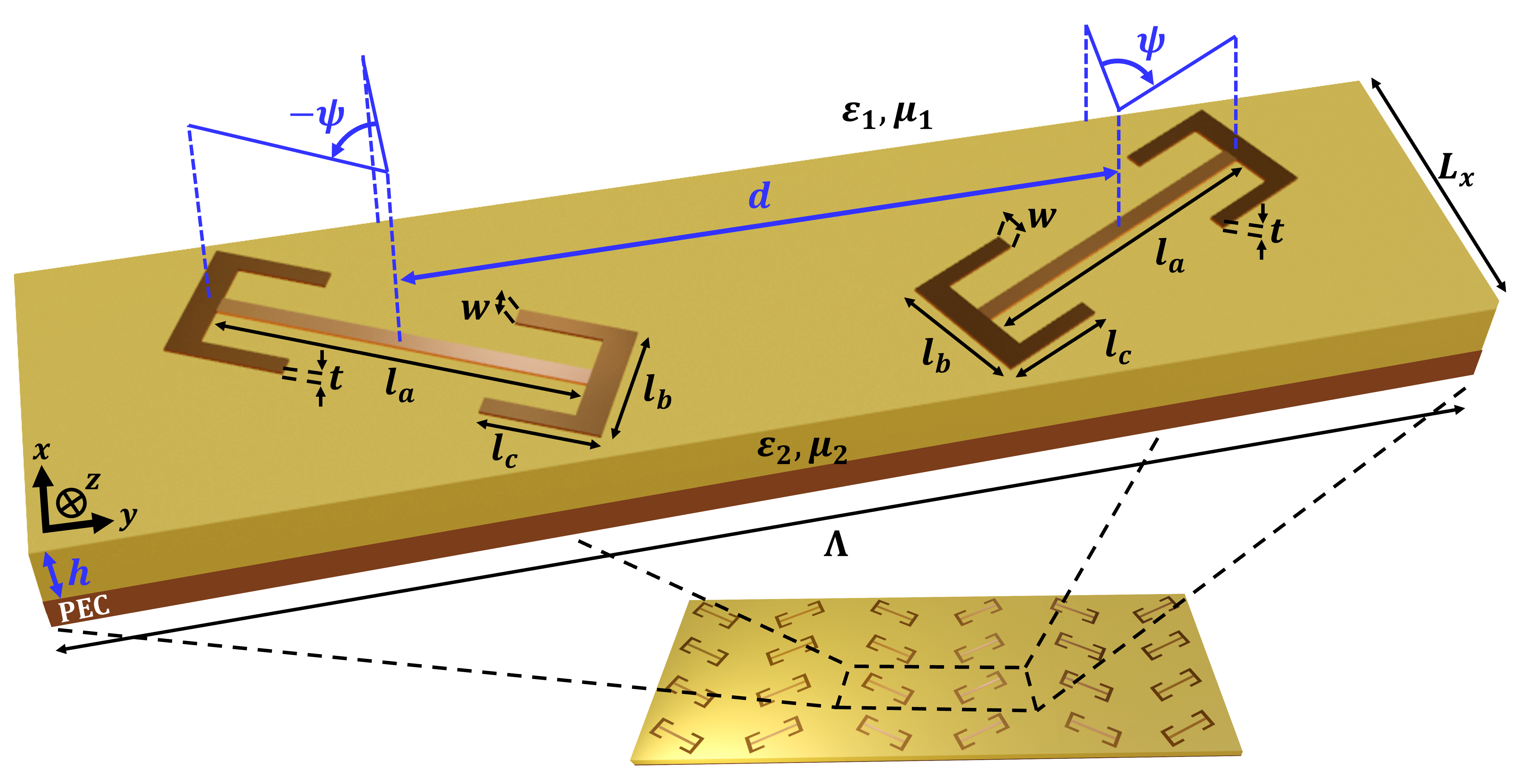}
\caption{\textcolor{black}{Symmetric private case of the (general) two-dipole scatterers per unit cell MG featured in Fig. \ref{Fig_config_and_general_UC}, corresponding to the configuration synthesized and analyzed in Section \ref{sec_Results_and_Discussion}. Herein, as opposed to Fig. \ref{Fig_config_and_general_UC}(b), the two \emph{identically} scaled meta-atoms are equidistant from the PEC ($h_1\!=\!h_2$) and are both located on the dielectric-air interface; they are inversely rotated with respect to one another ($\psi_1=-\psi_2$), featuring mirror symmetry within each unit cell.}}
\label{Fig_optimal_UC}
\end{figure}

Specifically for the chosen $L_x\!=\!\lambda/4$, the outlined synthesis procedure yields the following values for the designated functionality: $\tilde{I}_1\!=\!(0.5258+j0.3191)E^{\mathrm{TE}}_{\mathrm{in}}\Lambda/\eta_1$, $\tilde{I}_2\!=\!(0.2952+j0.5396)E^{\mathrm{TE}}_{\mathrm{in}}\Lambda/\eta_1$, $d\!=\!\Lambda/2\!=\!7.2088 \ \mathrm{mm}$, $\psi_1\!=\!-\psi_2\!=\!0.9588 \ \mathrm{rad}$ and $h_1\!=\!h_2\!=\!0.1364\lambda\!=\!2.0446 \ \mathrm{mm}$, corresponding to a coplanar equidistant arrangement of identical MAs with length $\ell$, all exhibiting a dipole moment of $|\tilde{I}_1\ell|\!=\!|\tilde{I}_2\ell|\!=\!0.615E^{\mathrm{TE}}_{\mathrm{in}}\Lambda\ell/\eta_1$ (up to a relative phase of $\angle\tilde{I}_1-\angle\tilde{I}_2\!=\!-30.07^\circ$) [Fig. \ref{Fig_optimal_UC}]. This symmetric nature of the obtained solution (which is not accidental), as well as its physical significance, will be discussed in Section \ref{subsec_Analytical_insights}.

Next, we validate the performance of this solution using a full-wave simulation in CST Microwave Studio. The MAs are implemented as realistic passive and lossy copper dog-bone shaped polarizable particles, positioned at the prescribed coordinates $(y,z)=(0,-h)$ and $(y,z)=(d,-h)$, $h\!\overset{\triangle}{=}\!h_{1}\!=\!h_{2}$, and tilted by $\psi\!\overset{\triangle}{=}\!\psi_{1}\!=\!-\psi_{2}$, with the precise values featured above, found by theory. \textcolor{black}{Their (conductor) width and thickness are set to $w\!=\!5\mathrm{mil}=127\mu\mathrm{m}$ and $t\!=\!18\mu\mathrm{m}$ respectively [Fig. \ref{Fig_optimal_UC}], chosen as to comply well with conventional PCB manufacturing resolution tolerances and standard (commercially available) electro-deposited copper foils.} By virtue of the MAs position and orientation symmetry and the received requirement for identical current amplitudes, it follows that their required physical sizes should also be identical, as elaborated in Section \ref{subsec_Analytical_insights}.

\textcolor{black}{To obtain optimal MA lengths $(l_a,l_b,l_c)$ relating to the dimensions in Fig. \ref{Fig_optimal_UC} – namely, the geometry that would yield an effective polarizability which would support the required induced currents once excited by the incident fields – an initial heuristic setting for their proportions was adopted, $(l_{a,i},l_{b,i},l_{c,i})=(0.245, 0.073, 0.067)\lambda$. Subsequently, these were multiplied by a \emph{common} scaling factor $\xi$ over which we performed a brief parametric sweep in CST under periodic boundary conditions to find the (optimal) lengths, seeking the highest coupling efficiency of the TE-polarized incident wave to the anomalous $m\!=\!-1$ TM-polarized FB mode. Executing this procedure for the MG parameters and functionality specified in this section resulted in such a valid solution ($\xi\!=\!0.84$), setting the MA dimensions to $(l_a,l_b,l_c)=(0.206,0.061,0.056)\lambda$.}

\begin{figure}[!b]
\centering
\includegraphics[width=\linewidth]{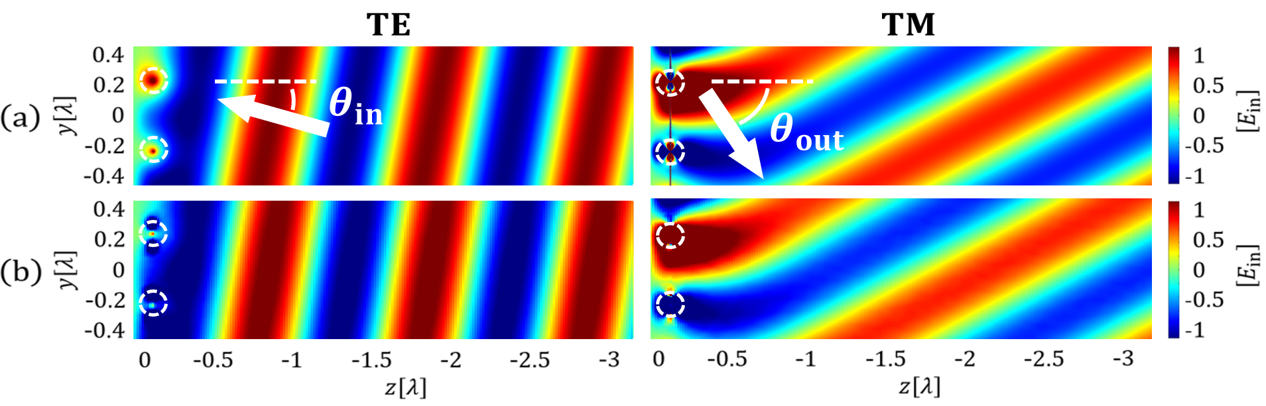}
\caption{Field distributions $\Re\{E_{x}^{\mathrm{tot}}(y,z)\}$ (left, TE polarization) and $\Re\{E_{y}^{\mathrm{tot}}(y,z)\}$ (right, TM polarization) across a single $y$-period, corresponding to the MG from Section \ref{subsec_Perfect_TE-to-TM}; designed to convert a TE polarized plane wave impinging upon the MG at $\theta_{\mathrm{in}}\!=\!10^\circ$, into a TM polarized plane wave anomalously reflected towards $\theta_{\mathrm{out}}\!=\!-\!60^\circ$. The white arrows indicate the direction of propagation, and the dashed white circles mark the two meta-atoms situated at $(y,z)\!=\!(\pm d/2,-h)$. Analytical predictions (a) following Eq. \eqref{subsec_Perfect_TE-to-TM} are compared with full-wave simulation of passive copper dog-bone shaped polarizable meta-atoms (b).}
\label{Fig_fields_plot}
\end{figure}

Fig. \ref{Fig_fields_plot} showcases the analytically predicted (a) and full-wave simulated (b) field distributions $\Re\{E_{x}^{\mathrm{tot}}(y,z)\}$ (left) and $\Re\{E_{y}^{\mathrm{tot}}(y,z)\}$ (right), corresponding to the TE  and TM  polarization components of the total electric field, respectively. As can be seen, these well agree, with some expected minor deviations near the MAs, due to their finite geometry not being accounted for in the analytical model \cite{Epstein2017UnveilingAnalysis, Rabinovich2018AnalyticalReflection}. Overall, the full-wave simulation indicates that the fraction of incident TE power from $\theta_\mathrm{in}$ converted to TM-polarized fields propagating along the anomalous reflection trajectory $\theta_\mathrm{out}$ is $\eta_c\!=\!96.5\%$, with as little as $0.68\%$ coupled to spurious TE or TM modes; and the rest $2.82\%$ is absorbed in the realistic copper conductors. The excellent correspondence between analytical theory and numerical simulation, as well as the high coupling efficiency obtained, verify the proposed concept and underlying model, yielding effective fabrication-ready anisotropic PCB MGs for simultaneous polarization conversion and beam deflection, as prescribed.

\textcolor{black}{We should emphasize that while in this section we demonstrated and verified our methodology for a single case study, with TE-to-TM polarization conversion alongside anomalous reflection from $\theta_\mathrm{in}\!=\!10^\circ$ to $\theta_\mathrm{out}\!=\!-60^\circ$, the scheme developed in Section \ref{sec_Theory} is entirely universal, and may effectively be employed for a broad range of input-output angle pairs, including shallow ones, as well as different incident wave polarization states. To further illustrate this universality, another case study is included in Appendix \ref{Appndx_D}, wherein a TE-polarized field from $\theta_\mathrm{in}\!=\!70^\circ$ is converted into a TM-polarized wave reflected towards $\theta_\mathrm{out}\!=\!-5^\circ$ with high efficiency, demonstrating an even more challenging (in terms of deflection angles) functionality. As also denoted in the beginning of this subsection, by virtue of the reciprocity theorem \cite{Tretyakov2003AnalyticalElectromagnetics}, these results imply the effectiveness of the devised technique also for TM-polarized incident fields, impinging upon the MG from either near-broadside or near-grazing angles.}

\textcolor{black}{Lastly, it is worthy of note that while the theoretical derivation guarantees that there should exist a passive and lossless scatterer which could support the prescribed functionality, it does not directly specify the geometry of such a MA. Consequently, although the representative cases considered in Section \ref{sec_Results_and_Discussion} and Appendix \ref{Appndx_D} (as well as others not shown) do demonstrate the substantial versatility of the proposed concept and synthesis methodology, it is difficult to assess accurately the limitations of the devised scheme in the context of practical realization in the microwave regime. A potential constraint is related to the lateral dimensions of the MA, which must remain deeply subwavelength to allow theoretical consideration as a rotated dipole line (i.e., having a dominant dipole moment), necessary to facilitate our analytical model and corresponding synthesis method (Section \ref{subsec_PolConv_&_AnoRef}). When large induced currents are required (e.g., when considering near-grazing angle of incidence/reflection, $|\theta_\mathrm{in/out}|\!\gtrsim\!75^\circ$), the MA polarizability should be significant as well, which generally requires larger dipolar scatterers (larger $l_a$). This may become a limiting factor in case one fails to sufficiently tune the resonant conditions via the loading arms ($l_b, l_c$). While a more rigorous treatment in the MA physical design is beyond the scope of this paper, such an extension, e.g., relying on \cite{Tretyakov2003AnalyticalElectromagnetics}, \cite{Shklarsh2024SemianalyticallyMetagratings}, will be incorporated into the methodology in future work.}

\subsection{Analytical insights} \label{subsec_Analytical_insights}

Reviewing the solution presented in Section \ref{subsec_Perfect_TE-to-TM} for the canonical TE-to-TM polarization conversion case (simultaneously with beam steering), yields an intriguing observation. Despite the random starting points for the inverse problem solution, enforcing the constraints of Section \ref{subsec_PolConv_&_AnoRef} to resolve the required MG parameters, the resultant \emph{optimal} set of solutions notably (and consistently) converged to specific configurations with distinct macro-period symmetry [Fig \ref{Fig_optimal_UC}]. In this clause, leveraging the high-fidelity analytical model and the element-sparse MG configuration, we shed light on this result. By devising closed-form expressions that tie the DOFs and the desired scattering functionality, we provide valuable physical insights regarding this optimal solution space and the role of the structural symmetry.

Recognizing that the optimal solutions of Section \ref{subsec_Perfect_TE-to-TM} feature coplanar and symmetric MA arrangements, we begin our analytical investigation by setting $h\!\overset{\triangle}{=}\!h_{1}\!=\!h_{2}$ (indicating that all MAs are located directly on the dielectric-air interface) and $\psi\!\overset{\triangle}{=}\!\psi_{1}\!=\!-\psi_{2}$, establishing mirror symmetry within each unit-cell [Fig. \ref{Fig_optimal_UC}]. Restricting the solution space as such is actually aligned with the number of constraints formulated in Section \ref{subsec_PolConv_&_AnoRef}, which, for the general (originally considered) configuration of Fig. \ref{Fig_config_and_general_UC} actually constitutes an underdetermined system of nonlinear equations \cite{Awange2018OverdeterminedSystems}; by defining these common $h$ and $\psi$, the number of DOFs is equalized with the number of constraints, leading to a balanced equation set.

Subsequently, we consider the same field excitation of Eq. \eqref{eq:E_in} (leading to the same external fields following Eqs. \eqref{eq:E^ext_1}-\eqref{eq:E^ext_2}), with an incidence angle $\theta_\mathrm{in}$ and a TE ($E_{\mathrm{in}}^{\mathrm{TM}}\!=\!0$) input polarization, and allow any user prescribed $\theta_\mathrm{out}$ angle which satisfies the the single-propagating-mode anomalous reflection constraint, as mentioned in section \ref{subsec_PolConv_&_AnoRef}.

Similar to Section \ref{subsec_Formulation}, we write the (modal) grid induced fields in region $1$ below the MG ($z<-h$)
\begin{equation}\label{eq:E^MG_1_opt}
\begin{aligned}
   & \boldsymbol{E}^{\mathrm{MG}}_{m,1}\left(y,z\right) \\
   & \ =\hat{x}\frac{1}{\Lambda}\tilde{I}_{m}^{\mathrm{TE}}\cos\psi B_{m,1}^{\mathrm{TE}}e^{j\beta_{m,1}z}e^{-jk_{t_{m}}y} \\
   & \ \ +\hat{y}\frac{1}{\Lambda}\tilde{I}_{m}^{\mathrm{TM}}\sin\psi B_{m,1}^{\mathrm{TM}}e^{j\beta_{m,1}z}e^{-jk_{t_{m}}y} \\
   & \ \ +\hat{z}\frac{1}{\Lambda}\tan\theta_{m,1}\ \tilde{I}_{m}^{\mathrm{TM}}\sin\psi B_{m,1}^{\mathrm{TM}}e^{j\beta_{m,1}z}e^{-jk_{t_{m}}y},
\end{aligned}
\end{equation}
and in region $2$ between the PEC and the dielectric-air interface ($-h<z<0$) \footnote{Herein, there is no distinction between regions $2^-$ and $2^+$ as $h_1\!=\!h_2$, hence the region between the PEC and the air-dielectric interface is simply denoted by $2$.}
\begin{equation}\label{eq:E^MG_2_opt}
\begin{aligned}
   & \boldsymbol{E}^{\mathrm{MG}}_{m,2}\left(y,z\right) \\
   & \ =-\hat{x}\frac{1}{\Lambda}\tilde{I}_{m}^{\mathrm{TE}}\cos\psi A_{m,2^{-}}^{\mathrm{TE}}\sin\left(\beta_{m,2}z\right)e^{-jk_{t_{m}}y} \\
   & \ \ -\hat{y}\frac{1}{\Lambda}\tilde{I}_{m}^{\mathrm{TM}}\sin\psi A_{m,2^{-}}^{\mathrm{TM}}\sin\left(\beta_{m,2}z\right)e^{-jk_{t_{m}}y} \\
   & \ \ +\hat{z}\frac{j}{\Lambda}\tan\theta_{m,2}\ \tilde{I}_{m}^{\mathrm{TM}}\sin\psi A_{m,2^{-}}^{\mathrm{TM}}\cos\left(\beta_{m,2}z\right)e^{-jk_{t_{m}}y}.
\end{aligned}
\end{equation} 
where, $A_{m,2}^{\mathrm{P}}$ and $B_{m,1}^{\mathrm{P}}$, the forward and backward amplitudes (in regions $2$ and $1$), respectively, are merely a private case of $A_{m,2}^{\mathrm{P},(1)}$ and $B_{m,1}^{\mathrm{P},(1)}$ featured in Eqs. \eqref{eq:E^MG_1}-\eqref{eq:E^MG_3} and defined explicitly in Eq. \eqref{eqB:A B}, for $h\!=\!h_1\!=\!h_2$. Herein, we utilize the definition of the effective generating secondary sources for the various scattered (polarization resolved) FB modes 
\begin{equation}\label{eq:I^TETM_m}
\begin{aligned}
   & \tilde{I}_{m}^{\mathrm{TE}}\!\overset{\triangle}{=}\!\tilde{I}_{1}\!+\!\tilde{I}_{2}e^{jk_{t_{m}}d}; \ \  \tilde{I}_{m}^{\mathrm{TM}}\!\overset{\triangle}{=}\!\tilde{I}_{1}\!-\!\tilde{I}_{2}e^{jk_{t_{m}}d},
\end{aligned}
\end{equation}
featuring electric current dimensions, contributed to by suitable superposition of the induced dipole moments per unit length Eq. \eqref{eq:I1I2}.

Altogether, the total fields are composed of the external ones [Eqs. \eqref{eq:E^ext_1}-\eqref{eq:E^ext_2}] and the secondary ones [Eqs. \eqref{eq:E^MG_1_opt}-\eqref{eq:E^MG_2_opt}], and by setting the period $\Lambda$ as in Section \ref{subsec_PolConv_&_AnoRef}, precisely two propagating modes (both TE and TM) are generated: the fundamental one made up of the specular reflection off the PEC and the $m\!=\!0$ FB mode, and the anomalous $m\!=\!-\!1$ FB mode. Just as in Eq. \eqref{eq:I1I2}, to eliminate the specular reflection for both TE and TM modes, we require that the corresponding terms in Eqs. \eqref{eq:E^ext_1} and \eqref{eq:E^MG_1_opt} would cancel each other. For the scenario considered herein, this condition sets the required \nth{0} order modal effective secondary sources according to the following closed-form expressions
\begin{equation}\label{eq:I^TETM_0}
\begin{aligned}
   & \tilde{I}_{0}^{\mathrm{TE}}=-\Lambda E_{\mathrm{in}}^{\mathrm{TE}}R_{0}^{\mathrm{TE}}\frac{e^{2j\beta_{0,1}h_{1}}}{B_{0,1}^{\mathrm{TE}}\cos\psi}; \ \ \ \tilde{I}_{0}^{\mathrm{TM}}\equiv0,
\end{aligned}
\end{equation}
thereby immediately setting the dipole moments per unit length via Eq. \eqref{eq:I^TETM_m}, namely
{\begin{equation}\label{eq:I_1,2 new}
\begin{aligned}
   &\tilde{I}_{1}=\frac{1}{2}\tilde{I}_{0}^{\mathrm{TE}}; \ \ \ \tilde{I}_{2}=\frac{1}{2}\tilde{I}_{0}^{\mathrm{TE}}e^{-jkd\sin\theta_{\mathrm{in}}}.
\end{aligned}
\end{equation}}

We take notice that this result implies that the two induced dipole moments per unit length should feature identical magnitudes $|\tilde{I}_{1}|\!=\!|\tilde{I}_{2}|$, while their phase differs by a factor $e^{-jkd\sin\theta_{\mathrm{in}}}$, coinciding with the effective current values presented in Section \ref{subsec_Perfect_TE-to-TM}. This observation is of utmost importance, unveiling a further degree of symmetry between each two MA pair: since the external fields [Eqs. \eqref{eq:E^ext_1}-\eqref{eq:E^ext_2}] exciting these induced currents also feature identical magnitude and $e^{-jkd\sin\theta_{\mathrm{in}}}$ phase difference when evaluated at the coordinates of the two MAs, the MA polarizability \cite{Tretyakov2003AnalyticalElectromagnetics} and consequently their geometry should be identical \emph{by definition}. Ergo, their dimensions are taken as identical in the aforementioned parametric sweep, reducing the solution space.

Once the specular reflection suppression is guaranteed via Eq. \eqref{eq:I^TETM_0}, we turn to realize the cross-polarization coupling to the prescribed anomalous $m\!=\!-1$ mode. To this end, we require \emph{perfect} anomalous reflection and polarization conversion, manifested as per  
\begin{equation}\label{eq:I^TETM_0,-1}
\begin{aligned}
   & \tilde{I}_{-1}^{\mathrm{TM}}\!=\!\tilde{I}_{0}^{\mathrm{TE}}; \ \ \ \tilde{I}_{-1}^{\mathrm{TE}}\!=\!\tilde{I}_{0}^{\mathrm{TM}}\!\equiv\!0,
\end{aligned}
\end{equation}
thereby suppressing the modal response of the undesired polarization and setting that of the desired one.

Interestingly, this requirement [Eq. \eqref{eq:I^TETM_0,-1}] has a very intuitive physical interpretation related to classical antenna array theory. When considering a single period as a subarray in the infinite MG, one may evaluate the corresponding subarray pattern using array factor (AF) considerations. In particular, considering the induced dipole moments $\tilde{I}_{1}$, $\tilde{I}_{2}$ on the two MAs, the distance between these secondary sources (acting as subarray elements), and their projection on the in plane axes (corresponding to their contribution to the polarized scattered fields) the individual TE and TM array factors (AFs) of each unit cell \cite{Balanis2016AntennaDesign} read
{\begin{equation}\label{eq:AF}
\begin{aligned}
   & AF^{\mathrm{TE}}\left(\theta\right)=\cos\psi\left(\tilde{I}_{1}+\tilde{I}_{2}e^{jkd\sin\theta}\right) \\
   & AF^{\mathrm{TM}}\left(\theta\right)=\sin\psi\left(\tilde{I}_{1}-\tilde{I}_{2}e^{jkd\sin\theta}\right).
\end{aligned}
\end{equation}}
Consequently, in order to suppress the specular reflection of the TM fields and ensure the coupling to the co-polar (TE-polarized) $m\!=\!-1$ FB mode, one should require that the corresponding array factors would have nulls in the associated scattering angles, namely $AF^{\mathrm{TE}}\left(\theta_{\mathrm{out}}\right)\!=\!AF^{\mathrm{TM}}\left(\theta_{\mathrm{in}}\right)\!=\!0$. The former requirement can be readily derived by virtue of Eq. \eqref{eq:I^TETM_0} (by plugging Eq. \eqref{eq:I_1,2 new} into Eq. \eqref{eq:AF} for $\theta\!=\!\theta_\mathrm{in}$), whereas the latter requirement is achieved by the same procedure (but for $\theta\!=\!\theta_\mathrm{out}$) and by additionally setting the interelement distance following $d\!=\!\Lambda/2\!=\!7.209 \ \mathrm{mm}$, yielding
{\begin{equation}\label{eq:AF_plugged_TE}
\begin{aligned}
   & AF^{\mathrm{TE}}\left(\theta_\mathrm{out}\right)\equiv0; \ \ \ AF^{\mathrm{TM}}\left(\theta_\mathrm{out}\right)=\tilde{I}_{0}^{\mathrm{TE}}\sin\psi,
\end{aligned}
\end{equation}}
altogether, well ascertaining Eq. \eqref{eq:I^TETM_0,-1} and elucidating the implementation of the equidistant MA arrangement. Thus, it can be seen that the suppression of spurious modes enforced by Eqs. \eqref{eq:I^TETM_0} and \eqref{eq:I^TETM_0,-1} (as a private case of Eqs. \eqref{eq:I1I2} and \eqref{eq:eta_-1} in Section \ref{subsec_PolConv_&_AnoRef}), is a direct consequence of antenna array theory and the symmetric configuration that was obtained from the inverse design solution.

Notwithstanding the suppression of these spurious modes, we still need to promote perfect anomalous reflection by requiring unitary coupling between the incident fundamental ($m=0$) TE mode and the reflected anomalous ($m\!=\!-1$) TM mode (as imposed by Eq. \eqref{eq:GlobPowerCons} of the general case). To facilitate such a complete power transfer, the MA tilt angle $\psi$ (responsible for cross-polarization coupling) should be tuned as to attain precise power balance between the TE and TM fields propagating in the two different angles, $\theta_\mathrm{in}$ and $\theta_\mathrm{out}$, respectively, accounting also for the wave impedance differences. For a given $h$, the net real power carried by the propagating waves, i.e. the $z$ component of Poynting’s vector across some plane below the MG (as in Eq. \eqref{eq:P_z^tot(mid)}), reads
{\begin{equation}\label{eq:Ptot_TE2TM}
\begin{aligned}
    & P_{z}^{\mathrm{tot}}(z<-h_{1}) \\ & \ =\frac{\Lambda}{2}\frac{1}{Z_{0,1}^{\mathrm{TE}}}\left|E_{\mathrm{in}}^{\mathrm{TE}}\right|^{2} \\
    & \ \ -\frac{\Lambda}{2}\frac{1}{Z_{-1,1}^{\mathrm{TM}}}\left|E_{\mathrm{in}}^{\mathrm{TE}}\right|^{2}\left|\frac{B_{-1,1}^{\mathrm{TM}}}{B_{0,1}^{\mathrm{TE}}}\right|^{2}\left|R_{0}^{\mathrm{TE}}\right|^{2}\tan^{2}\psi.
\end{aligned}
\end{equation}}
Enforcing perfect power transfer from the TE excitation channel to the anomalous TM one, requires this net power to vanish, leading to
{\begin{equation}\label{eq:psi_TE2TM}
\begin{aligned}
    & \tan\psi=\sqrt{\frac{Z_{-1,1}^{\mathrm{TM}}}{Z_{0,1}^{\mathrm{TE}}}}\left|B_{-1,1}^{\mathrm{TM}}\frac{R_{0}^{\mathrm{TE}}}{B_{0,1}^{\mathrm{TE}}}\right| \\
    & \ \ \ \ \ \ \ \ =\sqrt{\frac{Z_{-1,1}^{\mathrm{TM}}}{Z_{0,1}^{\mathrm{TE}}}}\left|\frac{Z_{-1,2}^{\mathrm{TM}}\frac{\tan\left(\beta_{-1,2}h\right)}{1+j\gamma_{-1}^{\mathrm{TM}}\tan\left(\beta_{-1,2}h\right)}}{Z_{0,2}^{\mathrm{TE}}\frac{\tan\left(\beta_{0,2}h\right)}{1-j\gamma_{0}^{\mathrm{TE}}\tan\left(\beta_{0,2}h\right)}}\right|,
\end{aligned}
\end{equation}}
resolving the MA orientation in closed-form (note, that the modal (backward) wave impedances ratios $\gamma_{m}^{\mathrm{TE}}\!=\!Z_{m,2}^{\mathrm{TE}} / Z_{m,1}^{\mathrm{TE}}$ and $\gamma_{m}^{\mathrm{TM}}\!=\!Z_{m,1}^{\mathrm{TM}} / Z_{m,2}^{\mathrm{TM}}$, are a generalization of the ones previously defined in Section \ref{subsec_Formulation}). Equation \eqref{eq:psi_TE2TM} again highlights the strength of the chosen formalism, providing a closed-form expression for the rotation angle, reflecting its the physical interpretation, controlling the coupling between the two polarization components (TE $m\!=\!0$ and TM $m\!=\!-1$) such that it will accommodate their desired amplitude ratio, all while considering the different modal wave impedances.

Finally, after setting the first 4 DOFs, i.e. the horizontal inter-element spacing $d\!=\!\Lambda/2$, the currents $\tilde{I}_{1},\tilde{I}_{2}$ via Eq. \eqref{eq:I_1,2 new} and the tilt angle $\psi$ via Eq. \eqref{eq:psi_TE2TM}, we utilize the remaining DOF $h$ such as to minimize the Ohmic losses within the copper traces (as was formulated in Eq. \eqref{eq:P_loss} of Section \ref{subsec_PolConv_&_AnoRef}. This step would enhance the overall efficiency of the device, even in the presence of realistic dissipation in the conductive MAs. Since, as was deduced in Eq. \eqref{eq:I_1,2 new}, the geometry of the two MAs is identical (up to their orientation), and since the conductor loss is proportional to the induced current squared \cite{Epstein2017UnveilingAnalysis}, the last design step would involve minimization of
{\begin{equation}\label{eq:I^2_TE2TM}
\begin{aligned}
&\left|\tilde{I}_{1}\right|^{2}\!+\!\left|\tilde{I}_{2}\right|^{2}\!=\!\left|\tilde{I}_{0}^{\mathrm{TE}}\right|^{2} \\
 & \ \ =\frac{\Lambda^{2}}{\cos^{2}\psi}\left|E_{\mathrm{in}}^{\mathrm{TE}}\right|^{2}\!\cdot\!\left|\frac{1-j\gamma_{0}^{\mathrm{TE}}\tan\left(\beta_{0,2}h\right)}{Z_{0,2}^{\mathrm{TE}}\tan\left(\beta_{0,2}h\right)}\right|^{2}.
\end{aligned}
\end{equation}}
Specifically, since the right-hand-side fraction includes an explicit dependency in $h$, we choose the MG thickness that leads to minimal induced dipole moments, aiming for the thinnest substrate that may accommodate this demand.

Ultimately, following this analytical solution-scheme, we stipulate exactly 5 synthesis constraints, used to resolve the 5 available DOFs in the MG configuration: Eq. \eqref{eq:I^TETM_0} sets the two dipole moments per unit length as to eliminate the TE and TM specular reflection; $d\!=\!\Lambda/2$ \emph{perfectly} suppresses the undesired TE polarization component of the anomalous reflection mode as indicated by Eqs. \eqref{eq:I^TETM_0,-1} and \eqref{eq:AF_plugged_TE}, by means of the MG modal response and its dual AF representation, respectively; Eq. \eqref{eq:psi_TE2TM} sets the elements relative tilt  angle as to balance the power coupled to the only two propagating modes, i.e. TE $m\!=\!0$ and TM $m\!=\!-\!1$; and via Eq. \eqref{eq:I^2_TE2TM} we find their smallest distance from the PEC which considerably minimizes the induced currents and hence the overall losses. Altogether, these constraints constitute as a well-determined system of nonlinear equations, manifesting our anisotropic PCB MG synthesis scheme for performing simultaneous beam steering and TE-to-TM polarization conversion.

Note that in comparison to analogous MS realizations featuring a similar symmetry \cite{Asadchy2016PerfectMetasurfaces, Yepes2021PerfectSlab}, the sheer simplicity of our proposed MG also provides physically-insightful closed-form expressions for the various design parameters explaining these symmetries, and yielding a compact synthesis framework.

As mentioned in Section \ref{subsec_Perfect_TE-to-TM}, the dual canonical case involving simultaneous TM-to-TE polarization conversion alongside anomalous reflection, is immediately retrieved from reciprocity. However, for the completeness of our discussion, we present in Appendix \ref{Appndx_C} the symmetric features and analytical derivation for this polarization conversion case, similar to the ones presented herein.

\subsection{Experiment} \label{subsec_Experiment}

After providing analytical insights regarding the MG design presented in Section \ref{subsec_Perfect_TE-to-TM}, we turn to verify experimentally the proposed concept. To this end, we strive to fabricate a prototype of this TE-to-TM $10^\circ$-to-$-60^\circ$ MG case study, building upon the corresponding analytical and full-wave simulation results [Fig. \ref{Fig_fields_plot}]. Considering now a more realistic version of the Rogers RO4350B laminate, featuring loss tangent of $\mathrm{tan}\delta\!=\!0.0037$ and an inherent anisotropic (uniaxial) permittivity of $\varepsilon_{zz}\!=\!3.66\varepsilon_0$ in the direction normal to the substrate and $\varepsilon_{xx}\!=\varepsilon_{yy}\!=\!3.75\varepsilon_0$ in the laminate plane \cite{ROGERS}, the optimal working point of the MG is slightly modified. While we retain the same MG thickness $h$ and MA tilt angle $\psi$ that were set by the semianlaytical design scheme for operation at $20$ GHz (Sections \ref{subsec_Perfect_TE-to-TM} and \ref{subsec_Analytical_insights}), the MA dimensions leading to the highest performance vary slightly (by ca. $2\%$) with respect to the ones found in Section \ref{subsec_Perfect_TE-to-TM} due to this laminate anisotropy and loss, reading $(l_a,l_b,l_c)=(0.201,0.06,0.055)\lambda$. Taking into account the additional dissipation in the dielectric somewhat reduces the overall cross-polar anomalous reflection efficiency, rating at $\eta_c\!=\!90.27\%$, with $0.98\%$ of the incident power being coupled to the residual modes and all other $8.75\%$ being absorbed in the copper traces \emph{and} in the lossy substrate \textcolor{black}{\footnote{\textcolor{black}{Note, that the simulated $\eta_c$ is the value of the relevant entry in the scattering matrix of the devised configuration, defined and simulated in CST. This scattering parameter, quantifying the scattered TM power coupled to the desired TM $m=-1$ FB mode relative to the overall incident TE power, is a part of the standard output by CST within such a Floquet port simulation.}}}.

\begin{figure}[!t]
\centering
\includegraphics[width=0.8\linewidth]{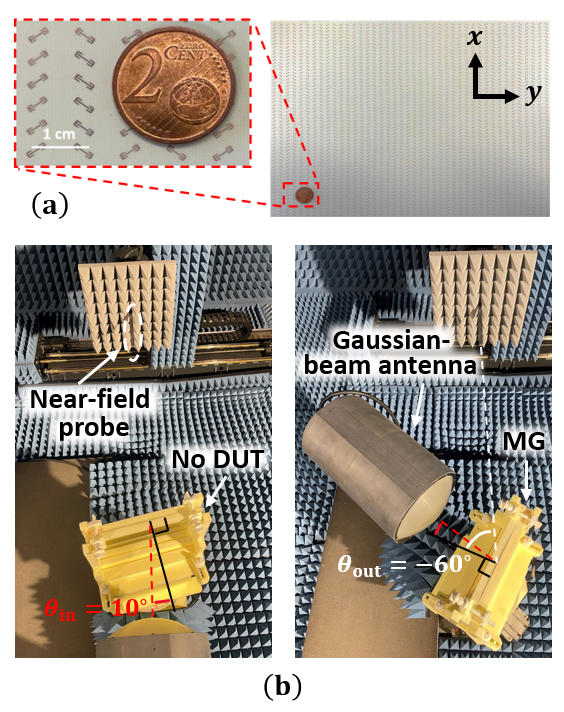}
\caption{(a) Front view of the fabricated MG prototype (right) corresponding to the symmetric two dogbone shaped meta-atoms per period design [Fig. \ref{Fig_optimal_UC}], and an inset (left). (b) Experiment setup inside the anechoic chamber.}
\label{Fig_prototype_and_exp_setup}
\end{figure}

\textcolor{black}{This fabrication-ready MG layout was realized as a
$9''\!\times\!12''$ ($22.86\mathrm{cm}\!\times\!30.48\mathrm{cm}$) board prototype using standard PCB manufacturing techniques, and is shown in Fig. \ref{Fig_prototype_and_exp_setup}(a). Due to limited availability of commercial laminate sizes and fabrication tolerances, the final thickness of the manufactured board was $77.5\mathrm{mil}$ ($1.969\mathrm{mm}$) instead of the required $80.5\mathrm{mil}$ ($2.0446\mathrm{mm}$).}

The MG device under test (DUT), i.e. the ultimately fabricated prototype, was characterized in an anechoic chamber at the Technion using a near-field measurement system (MVG/Orbit-FR Engineering Ltd., Emek Hefer, Israel); it was mounted on a foam holder and placed directly in front of a Gaussian beam antenna (Millitech, Inc., GOA-42-S000094) emitting a quasi-planar wavefront illumination, at a distance equal to its $196\mathrm{mm}\!\approx\!13\lambda$ focal length [Fig. \ref{Fig_prototype_and_exp_setup}(b)].

\begin{figure}[!t]
\centering
\includegraphics[width=\linewidth]{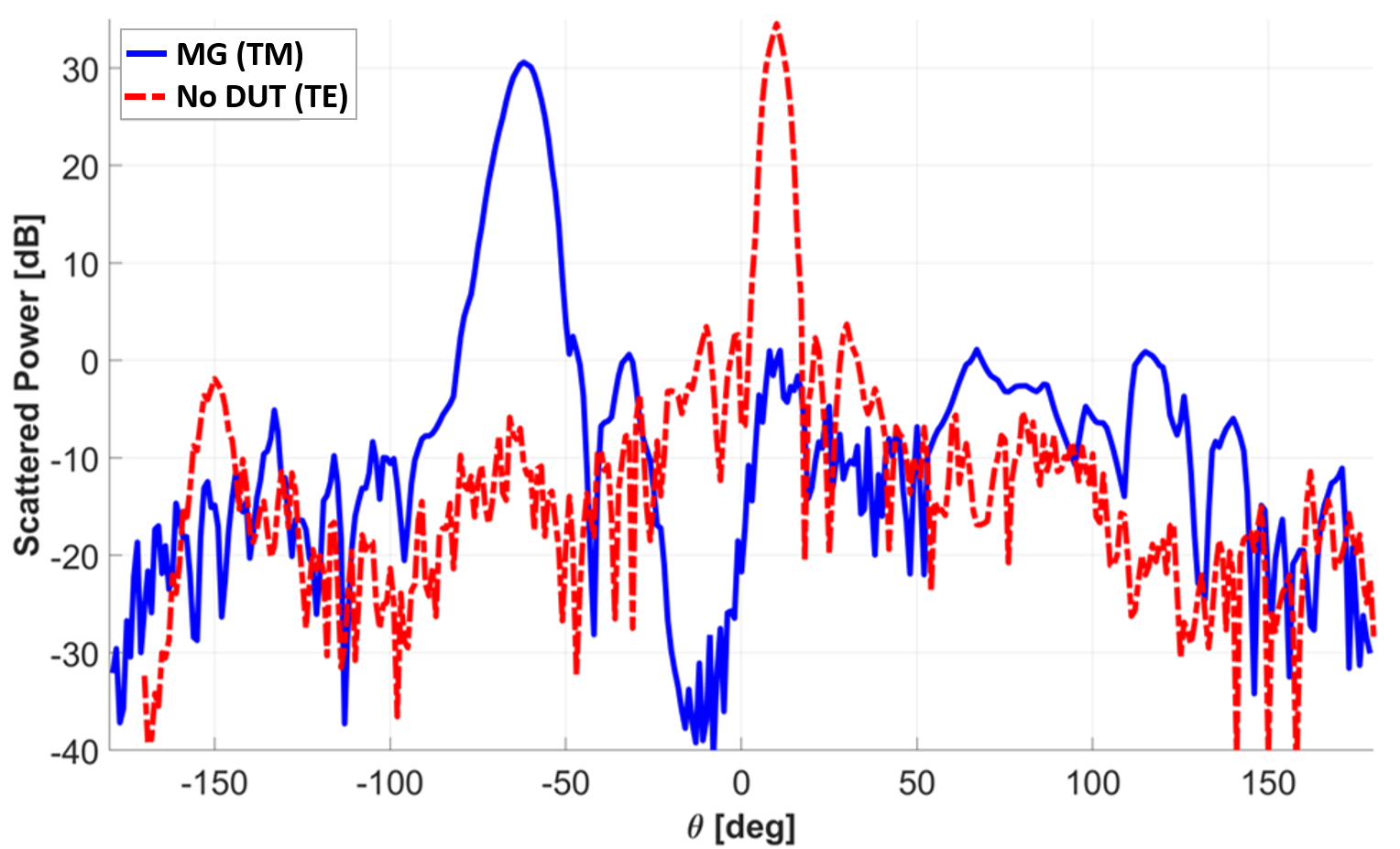}
\caption{Experimental scattering pattern of the illuminated prototype from Fig. \ref{Fig_prototype_and_exp_setup}(a), for TE-polarized illumination from a $\theta_\mathrm{in}\!=\!9^\circ$ incidence angle at the operating frequency of $20.1\mathrm{GHz}$. The scattered TM power reflected off the MG (solid blue), is compared to the reference input TE power radiation pattern recorded in the absence of the DUT (dash-dotted red).}
\label{Fig_radpat9}
\end{figure}

Figure \ref{Fig_radpat9} presents the recorded far-field radiation patterns of both the scattered TM-polarized reflection off the MG (solid blue) and the reference TE-polarized input illumination in the absence of the DUT (foam holder only, dash-dotted red), evaluated at the optimal measured working frequency of $20.1\mathrm{GHz}$. The minor $0.5\%$ spectral shift from the originally prescribed $20\mathrm{GHz}$ is ascribed to both the discrepancy in the dielectric thickness due to fabrication constraints mentioned earlier, and the manual alignment of the board with respect to the exciting antenna, which has an expected accuracy of $\pm1^\circ$. In particular, based on preliminary experimental studies we performed, it was concluded that the structure was ultimately illuminated from an incidence angle of $9^\circ$. Nonetheless, the recorded high TM-power peak at $-62^\circ$, corresponding to the theory predicted $-61.41^\circ$ (at these new frequency and illumination direction), indicates a very good redirection of the incoming TE-power towards the anomalous reflection trajectory, with the desired TM cross-polarization conversion.

\begin{figure}[!t]
\centering
\includegraphics[width=\linewidth]{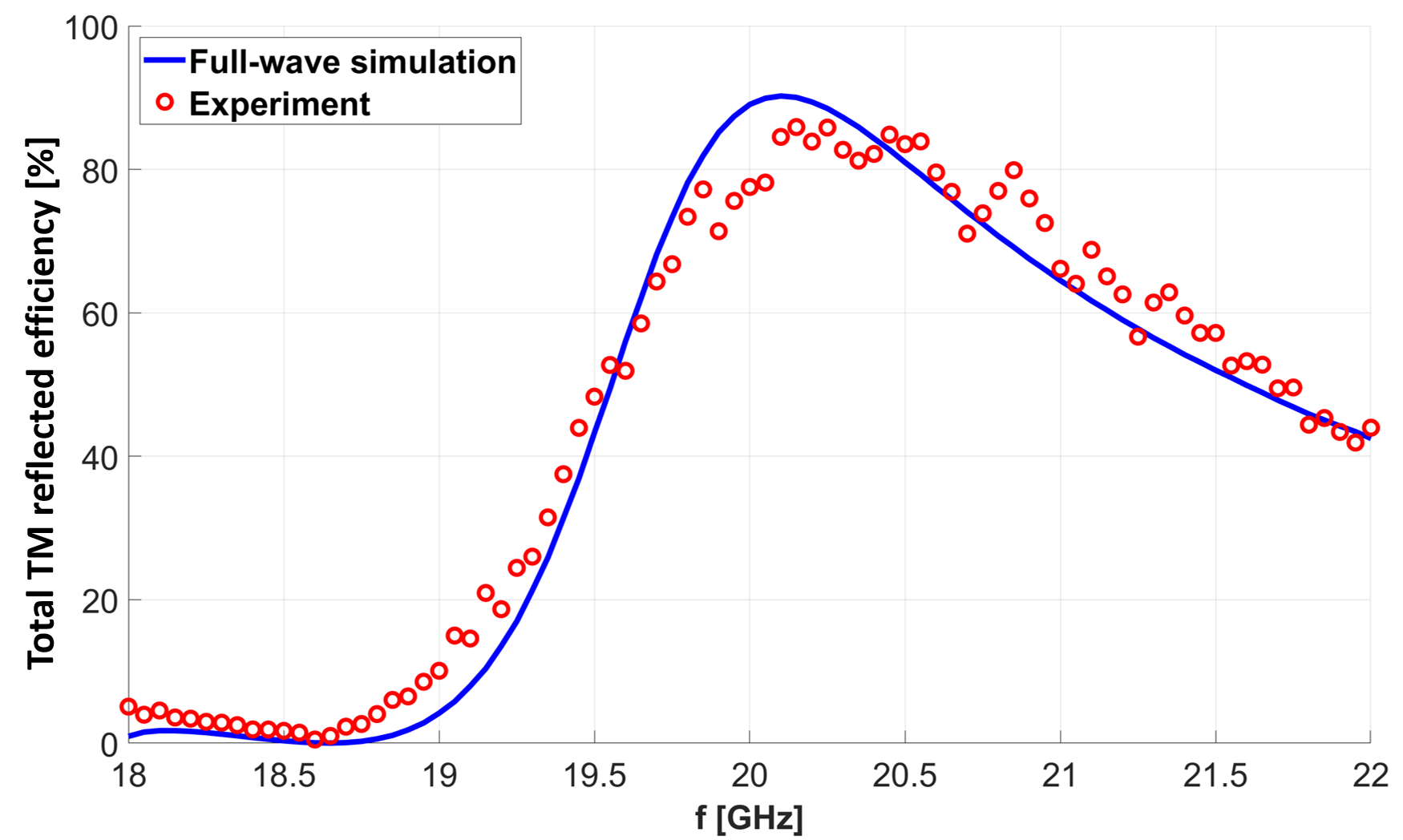}
\caption{Total TM reflected efficiency $\eta_c$ (frequency response) of the experimental illuminated prototype (red circles), compared to that of the full-wave simulation (solid blue), for illumination from a $\theta_\mathrm{in}\!=\!9^\circ$ incidence angle.}
\label{Fig_freqres9}
\end{figure}

With this regard, the experimentally measured and full-wave simulated reflected efficiencies, corresponding to the prototype MG at the actual experimental conditions, are presented in Fig. \ref{Fig_freqres9}. \textcolor{black}{Therein, the coupling efficiency is evaluated using \cite{Yashno2023BroadDiffusers}} 
\textcolor{black}{\begin{equation}\label{eq:eta_c}
\begin{aligned}
   & \eta_c=\frac{G^\mathrm{TM}_\mathrm{MG}(\theta_\mathrm{out})}{G^\mathrm{TE}_\mathrm{ref}(\theta_\mathrm{in})}\frac{\cos\theta_\mathrm{in}}{\cos\theta_\mathrm{out}},
\end{aligned}
\end{equation}}
\textcolor{black}{comparing the peak measured (TM) gain of the MG, $G_\mathrm{MG}(\theta_\mathrm{out})$, with the reference incident (TE) power in its absence, $G_\mathrm{ref}(\theta_\mathrm{in})$, evaluated directly from the measured (gain) radiation pattern \cite{Balanis2016AntennaDesign}. As discussed in \cite{Yashno2023BroadDiffusers}, the form of Eq. \ref{eq:eta_c} allows quantification of the power carried by the various scattered beams (corresponding to the various FB modes) for a given finite aperture, properly calibrating for the different effective aperture lengths for different observation angles via the factor ${\cos\theta_\mathrm{in}}/{\cos\theta_\mathrm{out}}$.} In reasonable agreement with the simulation predicted $\eta_c\!=\!90.23\%$, the experiment measured coupling efficiency peaking at the central frequency of $20.1\mathrm{GHz}$ rates at a high $\eta_c\!=\!85.9\%$, with a deviation of merely $4.33\%$. The received frequency response further validates the correspondence between the theoretical model and the recorded results.

Thus, utilizing both full-wave simulation and experimental measurement, the overall verification process showcased herein well verifies our theory and semianalytical synthesis framework, yielding highly efficient beam deflecting and polarization converting anisotropic MGs.

\section{Conclusion} \label{section_Conclusion}

To summarize, a systematic semi-analytical method for designing PCB-based metagratings, capable of \emph{simultaneous} anomalous beam deflection and polarization conversion, has been presented and verified through both full-wave simulation and experiment.

Our work began by proposing a model extension of the well studied canonical dipole-line. By allowing its periodically arranged meta-atom scatterers to be rotated in-plane by arbitrary rotation angles, we could allow this so called RDL to harvest, in principle, \emph{any} hybrid TE or TM polarization response. Incorporated into the detailed analytical MG synthesis scheme, this approach paves the way towards versatile polarization manipulating element-sparse MG devices, alternative to conventional polarization converting bulky MMs or challenging to design and implement MSs. 

Specifically, we investigated a two rotated meta-atoms per unit-cell PCB-implemented MG configuration, and provided a meticulous analytical description of how and why it is \emph{precisely} compatible for the canonical TE-to-TM polarization conversion task, simultaneously with anomalous reflection. By virtue of using element-sparse MGs, our analytical approach unveiled an intelligible connection between the various structural DOFs and the desired functionality, shedding light on the underlying physics at play; indicating that the ultimately received high-symmetry unit-cell [Fig. \ref{Fig_optimal_UC}] is in fact \emph{optimal} for orchestrating the constructive and destructive interference between all field components and polarizations. Basic antenna array theory concepts were utilized to elucidated the final configuration and its symmetrical geometry, leading, simply from symmetry considerations, to suppression of undesired dual polarized modes and full constructive interference in the anomalous direction.

To verify our theory, a specific TE-to-TM polarization conversion case study was considered, with prescribed incidence and output angles. Via a sweep in a full-wave solver, wherein a standard microwave-compatible PCB substrate was used, the analytically obtained solution was utilized to retrieve the realistic copper polarizable meta-atoms dimensions required to support the induced dipole moment. Yielding high efficiency, this provided complete fabrication ready specifications, which were utilized to fabricate a prototype of the MG, subsequently characterized in the lab. In line with the simulation results, the experimentally recorded high efficiency rate, even in the presence of realistic losses, serves an unambiguous validation of our theory and compellingly verifies the proposed synthesis scheme.

\textcolor{black}{Overall, the work fills critical gaps which currently exist in MS and MG analysis and synthesis. Regarding the latter, it proposes an architecture and lays out a systematic framework for extending the wavefront and polarization manipulation capabilities beyond single-polarized TE \cite{Epstein2017UnveilingAnalysis, Rabinovich2018AnalyticalReflection, Rabinovich2019ArbitraryMetagratings} or TM \cite{Memarian2017WidebandangleBS, Ra'di2018reconfigurable, Shklarsh2024SemianalyticallyMetagratings} MGs, allowing simultaneous polarization conversion and beam deflection with high efficiency in an appealing PCB platform \footnote{\textcolor{black}{Indeed, the experimentally evaluated efficiency of $85.9\%$ for the anomalous reflection and polarization conversion MG devised herein is comparable with other measured beam manipulating MG efficiencies of $\sim90\%$ reported in the past \cite{Rabinovich2019ExperimentalMetagratings}.}}. In the context of the former, it avoids the homogenization approximation hurdles via the formulated semianalytical model allowing swift effective practical designs for realizing this challenging functionality. Correspondingly, while some MSs implementing combined wave trajectory and polarization conversion have been theoretically envisioned before \footnote{\textcolor{black}{Being only theoretical and neglecting loss, it is somewhat difficult to compare the solutions in \cite{Asadchy2016PerfectMetasurfaces, Yepes2021PerfectSlab} to the designs presented herein. If losses are neglected for the design in Section \ref{subsec_Perfect_TE-to-TM}, the efficiency reaches $99.4\%$, almost identical to the optimal unitary MS solutions. For the simulated designs herein and in \cite{Yepes2021PerfectSlab}, one may also comment on the $90\%$ fractional bandwidth (with respect to the peak efficiency), finding a bandwidth of about $5\%$ for the results presented in Fig. \ref{Fig_freqres9} of Section \ref{subsec_Experiment} compared to about $2\%$ in \cite{Yepes2021PerfectSlab}.}}, the non-trivial need to translate the homogenized abstract GSTC-based highly anisotropic solutions into actual manufacturable layouts of numerous dense meta-atoms may require extensive full-wave optimization \cite{Yepes2021PerfectSlab} or end up with only a conceptual solution at hand \cite{Asadchy2016PerfectMetasurfaces}. In contrast, the MG approach devised herein directly yields fabrication-ready devices, with sparse element distributions that may also reduce manufacturing complexity. Importantly, utilizing the minimally required number of degrees of freedom in conjunction with the analytical model facilitates valuable analytical insights (Section \ref{subsec_Analytical_insights}), which shed light also on the symmetries apparent in related MS designs \cite{Yepes2021PerfectSlab}, in which such observations are sometimes more challenging to achieve.}

We are hopeful that the anisotropic MG topology introduced herein, and its accompanying straightforward and time-effective realization procedure, will inspire future work in this regard and lead to its integration in advanced vector beam forming systems. Specifically, adding additional MAs to each unit-cell would, in principle, enable more versatile functionalities at the output of the MG, similar to \cite{Rabinovich2019ArbitraryMetagratings}.

\appendix

\section{RDL electric field derivation} \label{Appndx_A}

For the benefit of the readers, we provide herein a detailed derivation of the transition between Eq. \eqref{eq:J^RDL} and Eq. \eqref{eq:E^RDL} in the main text, justifying the 2D analysis employed thereafter. To this end, we denote the LS and DL current densities per unit length ($L_x$), as per
\begin{equation}\label{eq:J(r)}
\begin{split}
    & J(\boldsymbol{r})\!\overset{\triangle}{=}\!\hat{x}\cdot\boldsymbol{J}^{\mathrm{LS}}(\boldsymbol{r})\!=\!\hat{y}\cdot\boldsymbol{J}^{\mathrm{DL}}(\boldsymbol{r}) \\
    & \ \ \ \ \ =\!I\ell\delta(y)\delta(z)\sum\nolimits _{n\in\mathbb{Z}}\delta\left(x\!-\!nL_{x}\right).
\end{split}
\end{equation}
By convoluting $J(\boldsymbol{r})$ with the (point source) Green's function $G(\boldsymbol{r})\!=\!e^{-jkr}\diagup4\pi r$, we derive the two components of the RDL magnetic vector potential, following \cite{Balanis2016AntennaDesign}

\begin{equation}\label{eq:A_e(r)_1}
\begin{split}
    & A_{e}(\rho) \\
    & \ \ \overset{\triangle}{=}\!\hat{x}\cdot\boldsymbol{A}_{e}^{\mathrm{LS}}(\boldsymbol{r})\!=\!\hat{y}\cdot\boldsymbol{A}_{e}^{\mathrm{DL}}(\boldsymbol{r}) \\
    & \ \ =\mu\intop_{\boldsymbol{r}'\in\mathbb{R}^{3}}d\boldsymbol{r}'\ J(\boldsymbol{r}-\boldsymbol{r}')G(\boldsymbol{r}) \\
    & \ \ =\frac{\mu}{4\pi}I\ell\sum_{n=-\infty}^{\infty}\frac{e^{-jk\sqrt{\left(x-nL_{x}\right)^{2}+\rho^{2}}}}{\sqrt{\left(x-nL_{x}\right)^{2}+\rho^{2}}} \\
    & \ \ \frac{\mu}{4\pi}I\ell\sum_{m\in\mathbb{Z}}\left[\ \intop_{l\in\mathbb{R}}dl\ \frac{e^{-jk\rho\sqrt{1+\left(\frac{x-lL_{x}}{\rho}\right)^{2}}}}{\rho\sqrt{1+\left(\frac{x-lL_{x}}{\rho}\right)^{2}}}e^{-j2\pi ml}\right],
\end{split}
\end{equation}
where $\rho\!=\!\sqrt{y^2+z^2}$ denotes the radial coordinate with respect to a cylindrical coordinate system centered around the $x$-axis, and in the last step we used the Poisson summation formula \cite{NIST:DLMF}. Now, by preforming the change of variables $\frac{l L_x-x}{\rho}\!=\!\sinh t$ and introducing the Mehler–Sonine integral \cite{NIST:DLMF}, we can evaluate the Fourier transform of the nontrivial integrand featured in Eq. \eqref{eq:A_e(r)_1}. Doing so, we arrive at
\begin{equation}\label{eq:A_e(r)_2}
\begin{split}
    & A_{e}(\rho,x)\!=\!-j\frac{\mu}{4}\tilde{I}\sum_{m\in\mathbb{Z}}H_{0}^{(2)}\left(k_{\rho,m}\rho\right)e^{-jk_{x,m}x},
\end{split}
\end{equation}
where, by virtue of the $x$-periodicity and the FB theorem, we define the wavenumber $k_{x,m}\!\overset{\triangle}{=}\!\frac{2\pi m}{L_{x}}\!=\!k\frac{\lambda m}{L_{x}}$ for propagation of the $m^\mathrm{th}$ FB mode along the $x$-axis ($m\!\in\!\mathbb{Z}$). Correspondingly, the transverse (radial) wavenumber is defined as $k_{\rho,m}\!=\!\sqrt{k^{2}-k_{x,m}^{2}}$, where $\mathfrak{Im}\left\{ k_{\rho,m}\right\} \!<\!0$ is chosen for physical outwards propagating waves.

Note, that the fundamental \nth{0} order mode has no dependence in $x$, and is propagating for any $L_x$. Thus, to eliminate the $x$-dependence altogether and constitute our 2D problem, we  suppress all other modes by taking $L_x<\lambda$, ensuring their evanescence. This leads to a simplified version of Eq. \eqref{eq:A_e(r)_2}, reading
\begin{equation}\label{eq:A_e(r)_3}
\begin{split}
    & A_{e}(\rho)\!=\!-\frac{j\mu}{4}\tilde{I}H_{0}^{(2)}(k\rho).
\end{split}
\end{equation}
From here, by carefully considering all field components and utilizing the well known relation between the magnetic vector potential and its induced electric field \cite{Balanis2016AntennaDesign}
\begin{equation}
\begin{split}
    & \boldsymbol{E}(\boldsymbol{r})\!=\!-j\omega\left[\boldsymbol{A}_{e}\left(\boldsymbol{r}\right)+\frac{1}{k^{2}}\nabla\left(\nabla\cdot\boldsymbol{A}_{e}\left(\boldsymbol{r}\right)\right)\right],
\end{split}
\end{equation}
one will arrive at Eq. \eqref{eq:E^RDL}.

It should be noted, that when considering $y$-periodic grids of such LSs and DLs (and generally, RDLs) in the context of MG or MS synthesis for instance, Eq. \eqref{eq:A_e(r)_3} for the magnetic vector potential holds valid only if the distance between the lines ($\Lambda$) is sufficient (comparable or above $L_x$). Moreover, to ensure the absolute evanescence of the higher order modes in practical designs, $L_x$ is taken as much smaller than the wavelength $\lambda$.

\section{Multiple reflection coefficients} \label{Appndx_B}

In this clause we derive explicit terms for the grids induced fields (TE and TM) forward and backward amplitudes $A^{\mathrm{P},(k)}_{m,i}$ and $B^{\mathrm{P},(k)}_{m,i}$, respectively, used in Eqs. \eqref{eq:E^MG_1}-\eqref{eq:E^MG_3} and featured in \ref{Fig_ab_coeffs} (Section \ref{subsec_Formulation}).

As in \cite{Rabinovich2019ArbitraryMetagratings}, the field in each region $i\!\in\!\left\{1,2^-,2^+\right\}$ generated by either RDL-grating source $k\!\in\!\left\{1,2\right\}$, can be written by virtue of the superposition principle as a sum of a forward- $A_{m,i}^{\mathrm{P},\left(k\right)}e^{-j\beta_{m,i}z}e^{-jk_{t_{m}}y}$ and backward- $B_{m,i}^{\mathrm{P},\left(k\right)}e^{j\beta_{m,i}z}e^{-jk_{t_{m}}y}$ traveling wave for either TE or TM polarization $\mathrm{P}$; all in all involving 24 unknowns. To explicitly obtain these amplitudes, tying the MG-generated fields to its DOFs, one must employ the relevant boundary conditions. These include the PEC boundary condition (vanishing of the tangential electric field at $z\!=\!0$), and the continuity of both the tangential electric and magnetic fields on the $z\!=\!-h_2$ and $z\!=\!-h_1$ interfaces, for sources $k\!=\!1$ and $k\!=\!2$, respectively. Moreover, due to the presence of the induced localized currents, i.e. sources $k\!=\!1$ and $k\!=\!2$ at $z\!=\!-h_1$ and $z\!=\!-h_2$, respectively, additional (source) conditions must be enforced, namely
\begin{equation} \label{eq:HyHx=JxJy}
\begin{split}
   & \left.H_{y}\right|_{h_{k}^{+}}-\left.H_{y}\right|_{h_{k}^{-}}\!=\!-\frac{1}{\Lambda}\tilde{I}_{k}\cos\psi_{k}e^{-jk_{t_{m}}\left(y-d_{k}\right)}\\
   & \left.H_{x}\right|_{h_{k}^{+}}-\left.H_{x}\right|_{h_{k}^{-}}\!=\!\frac{1}{\Lambda}\tilde{I}_{k}\sin\psi_{k}e^{-jk_{t_{m}}\left(y-d_{k}\right)}.
\end{split}
\end{equation}

When applying these conditions, half of the 24 unknowns either cancel out or immediately become redundant, following
{\begin{equation} \label{eq:ab=0}
\begin{split}
   & A_{m,1}^{\mathrm{P},(1)}=A_{m,1}^{\mathrm{P},(2)}\equiv0; \ \  A_{m,2^+}^{\mathrm{P},(2)}\!=\!-\!B_{m,2^+}^{\mathrm{P},(2)}; \\
   & A_{m,2^-}^{\mathrm{P},(1)}\!=\!A_{m,2^+}^{\mathrm{P},(1)}\!=\!-\!B_{m,2^+}^{\mathrm{P},(1)}\!=\!-\!B_{m,2^-}^{\mathrm{P},(1)}.
\end{split}
\end{equation}}
Thus, we are left with merely 6 amplitudes per polarization, namely 2 due to source $k\!=\!1$ [Fig. \ref{Fig_ab_coeffs}, blue-left], and 4 associated with source $k\!=\!2$ [Fig. \ref{Fig_ab_coeffs}, red-right], which can be directly found via the boundary conditions.

Alternatively, by utilizing the reflection and transmission coefficients from Eq. \eqref{eq:RT_bw}, the fields generated by the grids can be written in a recursive manner as in \cite{Rabinovich2019ArbitraryMetagratings} (therein only TE polarization was considered however). In this formulation, only the forward and backward amplitudes associated with the outwards-propagating waves \emph{directly} from the sources are involved, and all other amplitudes are written in terms of the reflection and transmission coefficients, or again, cancel out. Subsequently, employing the boundary conditions on the $z\!=\!-h_1$ and $z\!=\!-h_2$ interfaces for sources $k\!=\!1$ and $k\!=\!2$, respectively, leads to the explicit terms

\begin{equation}
\begin{split}\label{eqB:A B}
    & B_{m,1}^{\mathrm{P},\left(1\right)}=\frac{\tan\left(\beta_{m,2}h_{1}\right)e^{j\beta_{m,1}h_{1}}}{j\frac{1}{Z_{m,2}^{\mathrm{P}}}-\frac{1}{Z_{m,1}^{\mathrm{P}}}\tan\left(\beta_{m,2}h_{1}\right)} \\
    & A_{m,2^{-}}^{\mathrm{P},\left(1\right)}=\frac{e^{-j\beta_{m,1}h_{1}}}{\sin\left(\beta_{m,2}h_{1}\right)}B_{m,1}^{\mathrm{P},\left(1\right)} \\
    & B_{m,2^{-}}^{\mathrm{P},\left(2\right)}=Z_{m,2}^{\mathrm{P}}\tan\left(\beta_{m,2}h_{2}\right)e^{j\beta_{m,2}h_{2}}e^{j\left(\beta_{m,1}\!-\!\beta_{m,2}\right)h_{1}} \\
    & \ \ \ \ \ \ \ \ \ \ \ \cdot\left[j\left(1\!+\!\hat{R}_{m}^{\mathrm{P}}e^{2j\beta_{m,2}\left(h_{2}\!-\!h_{1}\right)}\right)\right. \\
    & \ \ \ \ \ \ \ \ \ \ \ \ -\left.\left(1\!-\!\hat{R}_{m}^{\mathrm{P}}e^{2j\beta_{m,2}\left(h_{2}\!-\!h_{1}\right)}\right)\tan\left(\beta_{m,2}h_{2}\right)\right] \\
    & A_{m,2^{+}}^{\mathrm{P},\left(2\right)}\!=\!\frac{1\!+\!\hat{R}_{m}^{\mathrm{P}}e^{2j\beta_{m,2}\left(h_{2}\!-\!h_{1}\right)}}{\sin\left(\beta_{m,2}h_{2}\right)}e^{j\left(\beta_{m,2}\!-\!\beta_{m,1}\right)h_{1}}B_{m,2^{-}}^{\mathrm{P},\left(2\right)}, \\
\end{split}
\end{equation}
where the (TE and TM) wave impedances $Z_{m,i}^{\mathrm{P}}$ and the reflection and transmission coefficients $\hat{R}_m^\mathrm{P}$ and $\hat{T}_m^\mathrm{P}$, are defined as in Section \ref{subsec_Formulation}.
\newpage

\section{Perfect TM-to-TE polarization conversion and anomalous reflection} \label{Appndx_C}

Similar to the TE to TM polarization conversion case (simultaneously with beam steering) showcased in Section \ref{subsec_Analytical_insights}, the dual TM to TE task also features similar analytical insights. Herein, we briefly present the analogous synthesis equations for this case, for given $\theta_\mathrm{in}$, $\theta_\mathrm{out}$, and a TM-polarized incident field.

By virtue of suppressing the specular reflection as in Eq. \eqref{eq:I^TETM_0}, the effective (\nth{0} order) secondary sources read
\begin{equation}\label{eqC:I^TETM_0}
\begin{aligned}
   & \tilde{I}_{0}^{\mathrm{TM}}=-\Lambda E_{\mathrm{in}}^{\mathrm{TM}}\frac{R_{0}^{\mathrm{TM}}e^{2j\beta_{0,1}h}}{B_{0,1}^{\mathrm{TM}}\sin\psi}\cos\theta_{\mathrm{in}}; \ \tilde{I}_{0}^{\mathrm{TE}}\equiv0,
\end{aligned}
\end{equation}
which set the currents following Eq. \eqref{eq:I^TETM_m}
{\begin{equation}\label{eqC:I_1,2 new}
\begin{aligned}
   &\tilde{I}_{1}=\frac{1}{2}\tilde{I}_{0}^{\mathrm{TM}}; \ \tilde{I}_{2}=-\frac{1}{2}\tilde{I}_{0}^{\mathrm{TM}}e^{-jkd\sin\theta_{\mathrm{in}}}.
\end{aligned}
\end{equation}}

Once more, we \emph{perfectly} suppress the polarization component of the undesired anomalous reflection (TM in this case), by simply setting $d\!=\!\Lambda/2$, thusly receiving
\begin{equation}\label{eqC:I^TETM_0,-1}
\begin{aligned}
   & \tilde{I}_{-1}^{\mathrm{TE}}\!=\!\tilde{I}_{0}^{\mathrm{TM}};\ \tilde{I}_{-1}^{\mathrm{TM}}\!=\!\tilde{I}_{0}^{\mathrm{TE}}\!\equiv\!0,
\end{aligned}
\end{equation}
which, in AF terms following Eq. \eqref{eq:AF}, translates to
{\begin{equation}\label{eq:AF_plugged_TM}
\begin{aligned}
    & AF^{\mathrm{TM}}\left(\theta\right)=\tilde{I}_{0}^{\mathrm{TM}}\sin\psi;\ AF^{\mathrm{TE}}\left(\theta\right)=0.
\end{aligned}
\end{equation}}

Next, to funnel all the incident TM power to the TE anomalous reflection mode (as equivalently done in Eqs. \eqref{eq:Ptot_TE2TM} and \eqref{eq:psi_TE2TM}), we equate the net real power beyond the device, reading
{\begin{equation}\label{eq:Ptot_TM2TE}
\begin{aligned}
    & P_{z}^{\mathrm{tot}}(z<-h_{1}) \\ & \ =\frac{\Lambda}{2}\frac{\cos^{2}\theta_{\mathrm{in}}}{Z_{0,1}^{\mathrm{TM}}}\left|E_{\mathrm{in}}^{\mathrm{TM}}\right|^{2} \\
    & \ \  -\frac{\Lambda}{2}\frac{\cos^{2}\theta_{\mathrm{in}}}{Z_{-1,1}^{\mathrm{TE}}}\left|E_{\mathrm{in}}^{\mathrm{TM}}\right|^{2}\left|R_{0}^{\mathrm{TM}}\right|^{2}\left|\frac{B_{-1,1}^{\mathrm{TE}}}{B_{0,1}^{\mathrm{TM}}}\right|^{2}\cot^{2}\psi
\end{aligned}
\end{equation}}
to zero, and obtain a closed-form term for the tilt angle $\psi$ relying only on $h$, namely
{\begin{equation}\label{eq:psi_TM2TE}
\begin{aligned}
    & \tan\psi=\sqrt{\frac{Z_{0,1}^{\mathrm{TM}}}{Z_{-1,1}^{\mathrm{TE}}}}\left|B_{-1,1}^{\mathrm{TE}}\frac{R_{0}^{\mathrm{TM}}}{B_{0,1}^{\mathrm{TM}}}\right| \\
    & \ \ \ \ \ \ \ \ =\sqrt{\frac{Z_{0,1}^{\mathrm{TM}}}{Z_{-1,1}^{\mathrm{TM}}}}\left|\frac{Z_{-1,2}^{\mathrm{TE}}\frac{\tan\left(\beta_{-1,2}h\right)}{1+j\gamma_{-1}^{\mathrm{TE}}\tan\left(\beta_{-1,2}h\right)}}{Z_{0,2}^{\mathrm{TM}}\frac{\tan\left(\beta_{0,2}h\right)}{1-j\gamma_{0}^{\mathrm{TM}}\tan\left(\beta_{0,2}h\right)}}\right|.
\end{aligned}
\end{equation}}

Lastly, by considering the overall currents reading
{\begin{equation}\label{eq:I^2_TM2TE}
\begin{aligned}
& \left|\tilde{I}_{1}\right|^{2}\!+\!\left|\tilde{I}_{2}\right|^{2}\!=\!\left|\tilde{I}_{0}^{\mathrm{TE}}\right|^{2} \\
& \ \ = \frac{\Lambda^{2}}{\sin^{2}\psi}\left|E_{\mathrm{in}}^{\mathrm{TM}}\right|^{2}\cos^{2}\theta_{\mathrm{in}}\left|\frac{1-j\gamma_{0}^{\mathrm{TM}}\tan\left(\beta_{0,2}h\right)}{Z_{0,2}^{\mathrm{TM}}\tan\left(\beta_{0,2}h\right)}\right|^{2},
\end{aligned}
\end{equation}}
and minimizing them to reduce the Ohmic losses, we set the final DOF $h$, all in all arriving at a complete set of values for all the DOFs.

\textcolor{black}{\section{\textcolor{black}{Wide-angle simultaneous polarization conversion and anomalous reflection with shallow-angle incidence}} \label{Appndx_D}}

\textcolor{black}{To further exemplify the versatility of our synthesis framework, we showcase an additional second case study to the one presented in Section \ref{subsec_Perfect_TE-to-TM}, considering both a relatively shallower incidence angle and a more acute output angle. Herein, a TE-polarized wave impinging at a $\theta_\mathrm{in}\!=\!70^\circ$ incidence angle, is reflected as a TM-polarized wave, exiting at an anomalous $\theta_\mathrm{out}\!=\!-5^\circ$ output angle. The analytically-derived DOFs values for this specific case read: $\tilde{I}_1\!=\!(0.0487\!+\!j0.0378)E^{\mathrm{TE}}_{\mathrm{in}}\Lambda/\eta_1$, $\tilde{I}_2\!=\!(\!-\!0.037\!-\!j0.0493)E^{\mathrm{TE}}_{\mathrm{in}}\Lambda/\eta_1$, $d\!=\!\Lambda/2\!=\!7.2088\mathrm{mm}$, $\psi_1\!=\!-\!\psi_2\!=\!1.0305 \ \mathrm{rad}$ and $h_1\!=\!h_2\!=\!0.1428\lambda\!=\!2.1403\mathrm{mm}$.}

\begin{figure}[!b]
\centering
\includegraphics[width=\linewidth]{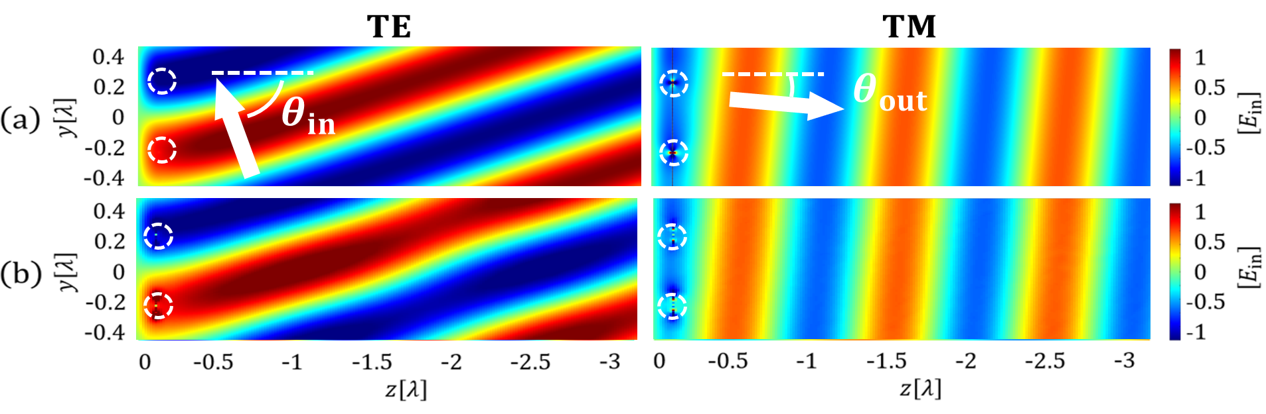}
\caption{\textcolor{black}{Field distributions $\Re\{E_{x}^{\mathrm{tot}}(y,z)\}$ (left, TE polarization) and $\Re\{E_{y}^{\mathrm{tot}}(y,z)\}$ (right, TM polarization) across a single $y$-period, corresponding to the MG from Appendix \ref{Appndx_D}; designed to convert a TE polarized plane wave impinging upon the MG at $\theta_{\mathrm{in}}\!=\!70^\circ$, into a TM polarized plane wave anomalously reflected towards $\theta_{\mathrm{out}}\!=\!-\!5^\circ$. All other parameters are identical to the ones listed in the caption of Fig. \ref{Fig_fields_plot}.}}
\label{Fig_fields_appendix}
\end{figure}

\textcolor{black}{The theoretically predicted $100\%$ perfect coupling efficiency, is met with a high $\eta_c\!=\!93.29\%$ received from a full-wave simulation including realistic copper MAs and a lossy and anisotropic RO4350B substrate, as used in Section \ref{subsec_Perfect_TE-to-TM}. Moreover, merely $0.73\%$ of the incident power is coupled to other spurious modes, implying that the residual $5.98\%$ is absorbed as loss by the device. Note, that optimal MA dimensions were again found through a similar parametric sweep procedure as described in Section \ref{subsec_Perfect_TE-to-TM}, resulting in $(l_a,l_b,l_c)=(0.164,0.049,0.048)\lambda$; all other parameters remained unchanged, as listed for the simulation in Section \ref{subsec_Perfect_TE-to-TM}. In analogy to Fig. \ref{Fig_fields_plot}, Fig. \ref{Fig_fields_appendix} showcases the fields distributions visualizing the good correspondence between theory and simulation, once more verifying our semianalytical synthesis scheme.} 

\textcolor{black}{We should note that due to the reciprocity theorem, the case studies presented in Section \ref{subsec_Perfect_TE-to-TM} and Appendix \ref{Appndx_D} demonstrate, in fact, also the capability of the proposed scheme to devise TM-to-TE polarization converting anomalous reflectors, for both shallow and near-normal angles of incidence, respectively. In particular, reciprocity dictates that $\eta_c$ should remain identical when replacing the excitation and observation probes, i.e., when illuminating the MG by a TM-polarized field incident from what was defined in Section \ref{subsec_Perfect_TE-to-TM} and Appendix \ref{Appndx_D} as $\theta_\mathrm{out}$ and considering the coupling towards $\theta_\mathrm{in}$. Indeed, when simulating the MG designed in Section \ref{subsec_Perfect_TE-to-TM} (unchanged) with an illumination of a TM-polarized wave from $\theta_\mathrm{in}=60^\circ$, the coupling to $\theta_\mathrm{out}=-10^\circ$ as retrieved from the simulation is $96.8\%$, with $0.64\%$ coupled to spurious modes and $2.56\%$ dissipated as loss (vs. the original efficiency rates $96.5\%$, $0.68\%$, and $2.82\%$, respectively). Similarly, when simulating the MG designed in Appendix \ref{Appndx_D} (unchanged) with an illumination of a TM-polarized wave from $\theta_\mathrm{in}=5^\circ$,  the coupling to $\theta_\mathrm{out}=-70^\circ$ as retrieved from the simulation is $93.53\%$, with $0.62\%$ coupled to spurious modes and $5.68\%$ dissipated as loss (vs. the original efficiency rates $93.29\%$, $0.72\%$, and $5.75\%$, respectively). The minor deviations of $\sim0.05\%$ from the reciprocity theorem predictions are attributed to numerical errors within the full-wave solver.}

\bibliographystyle{IEEEtran}
\bibliography{references}

\begin{thebibliography}{10}
\providecommand{\url}[1]{#1}
\csname url@samestyle\endcsname
\providecommand{\newblock}{\relax}
\providecommand{\bibinfo}[2]{#2}
\providecommand{\BIBentrySTDinterwordspacing}{\spaceskip=0pt\relax}
\providecommand{\BIBentryALTinterwordstretchfactor}{4}
\providecommand{\BIBentryALTinterwordspacing}{\spaceskip=\fontdimen2\font plus
\BIBentryALTinterwordstretchfactor\fontdimen3\font minus \fontdimen4\font\relax}
\providecommand{\BIBforeignlanguage}[2]{{%
\expandafter\ifx\csname l@#1\endcsname\relax
\typeout{** WARNING: IEEEtran.bst: No hyphenation pattern has been}%
\typeout{** loaded for the language `#1'. Using the pattern for}%
\typeout{** the default language instead.}%
\else
\language=\csname l@#1\endcsname
\fi
#2}}
\providecommand{\BIBdecl}{\relax}
\BIBdecl

\bibitem{GlybovskiMetasurfacesVisible}
S.~B. Glybovski, S.~A. Tretyakov, P.~A. Belov, Y.~S. Kivshar, and C.~R. Simovski, ``{Metasurfaces: From Microwaves to Visible},'' \emph{Phys. Rep.}, vol. 634, pp. 1--72, 2016.

\bibitem{Raadi2022MetagratingsManipulation}
Y.~Ra'di and A.~Al{\`u}, ``{Metagratings for Efficient Wavefront Manipulation},'' \emph{IEEE Photon. J.}, vol.~14, no.~1, pp. 1--13, 2021.

\bibitem{Iyer2020MetamaterialsDirections}
A.~K. Iyer, A.~Al{\`u}, and A.~Epstein, ``{Metamaterials and Metasurfaces — Historical Context, Recent Advances, and Future Directions},'' \emph{IEEE Trans. Antennas Propag.}, vol.~68, no.~3, pp. 1223--1231, 2020.

\bibitem{Tretyakov2003AnalyticalElectromagnetics}
S.~Tretyakov, \emph{{Analytical Modeling in Applied Electromagnetics}}.\hskip 1em plus 0.5em minus 0.4em\relax Artech House, 2003.

\bibitem{Kuester2003AveragedMetafilm}
E.~F. Kuester, M.~A. Mohamed, M.~Piket-May, and C.~L. Holloway, ``{Averaged Transition Conditions for Electromagnetic Fields at a Metafilm},'' \emph{IEEE Trans. Antennas Propag.}, vol.~51, no.~10, pp. 2641--2651, 2003.

\bibitem{Epstein2016HuygensApplications}
A.~Epstein and G.~V. Eleftheriades, ``{Huygens’ Metasurfaces via the Equivalence Principle: Design and Applications},'' \emph{J. Opt. Soc. Am. B}, vol.~33, no.~2, pp. A31--A50, 2016.

\bibitem{Radi2017MetagratingsControl}
Y.~Ra’di, D.~L. Sounas, and A.~Al{\`u}, ``{Metagratings: Beyond the Limits of Graded Metasurfaces for Wave Front Control},'' \emph{Phys. Rev. Lett.}, vol. 119, no.~6, p. 067404, 2017.

\bibitem{Epstein2017UnveilingAnalysis}
A.~Epstein and O.~Rabinovich, ``{Unveiling the Properties of Metagratings via a Detailed Analytical Model for Synthesis and Analysis},'' \emph{Phys. Rev. Appl.}, vol.~8, no.~5, p. 054037, 2017.

\bibitem{Rabinovich2018AnalyticalReflection}
O.~Rabinovich and A.~Epstein, ``{Analytical Design of Printed Circuit Board (PCB) Metagratings for Perfect Anomalous Reflection},'' \emph{IEEE Trans. Antennas Propag.}, vol.~66, no.~8, pp. 4086--4095, 2018.

\bibitem{Memarian2017WidebandangleBS}
M.~Memarian, X.~Li, Y.~Morimoto, and T.~Itoh, ``{Wide-band/angle Blazed Surfaces using Multiple Coupled Blazing Resonances},'' \emph{Sci. Rep.}, vol.~7, no.~1, pp. 1--12, 2017.

\bibitem{Sell2017Large-AngleGeometries}
D.~Sell, J.~Yang, S.~Doshay, R.~Yang, and J.~A. Fan, ``{Large-angle, Multifunctional Metagratings Based on Freeform Multimode Geometries},'' \emph{Nano Lett.}, vol.~17, no.~6, pp. 3752--3757, 2017.

\bibitem{Wong2018PerfectMetasurface}
A.~M. Wong and G.~V. Eleftheriades, ``{Perfect Anomalous Reflection with a Bipartite Huygens’ Metasurface},'' \emph{Phys. Rev. X.}, vol.~8, no.~1, p. 011036, 2018.

\bibitem{Yang2018FreeformSteering}
J.~Yang, D.~Sell, and J.~A. Fan, ``{Freeform Mmetagratings Based on Complex Light Scattering Dynamics for Extreme, High Efficiency Beam Steering},'' \emph{Annalen der Physik}, vol. 530, no.~1, p. 1700302, 2018.

\bibitem{Dong2020EfficientMetagratings}
X.~Dong, J.~Cheng, F.~Fan, X.~Wang, and S.~Chang, ``Efficient wide-band large-angle refraction and splitting of a terahertz beam by low-index 3d-printed bilayer metagratings,'' \emph{Phys. Rev. Appl.}, vol.~14, p. 014064, 7 2020.

\bibitem{Rabinovich2019ArbitraryMetagratings}
O.~Rabinovich and A.~Epstein, ``{Arbitrary Diffraction Engineering with Multilayered Multielement Metagratings},'' \emph{IEEE Trans. Antennas Propag.}, vol.~68, no.~5, pp. 1553--1568, 2019.

\bibitem{Popov2018ControllingMetagratings}
V.~Popov, F.~Boust, and S.~N. Burokur, ``{Controlling Diffraction Patterns with Metagratings},'' \emph{Phys. Rev. Appl.}, vol.~10, no.~1, p. 011002, 2018.

\bibitem{Popov2019ConstructingFreedom}
------, ``Constructing the near field and far field with reactive metagratings: Study on the degrees of freedom,'' \emph{Phys. Rev. Appl.}, vol.~11, no.~2, p. 024074, 2019.

\bibitem{Rabinovich2019ExperimentalMetagratings}
O.~Rabinovich, I.~Kaplon, J.~Reis, and A.~Epstein, ``Experimental demonstration and in-depth investigation of analytically designed anomalous reflection metagratings,'' \emph{Phys. Rev. B}, vol.~99, p. 125101, Mar 2019.

\bibitem{Xu2021DualWires}
G.~Xu, S.~V. Hum, and G.~V. Eleftheriades, ``{Dual-Band Reflective Metagratings With Interleaved Meta-Wires},'' \emph{IEEE Trans. Antennas Propag.}, vol.~69, no.~4, pp. 2181--2193, 2021.

\bibitem{Rabinovich2019PrintedExperiment}
O.~Rabinovich, Y.~Komarovsky, D.~Dikarov, and A.~Epstein, ``Printed circuit board (pcb) metagratings for perfect anomalous refraction: Theory and experiment,'' in \emph{ICEAA 2019}, 2019, pp. 1070--1070.

\bibitem{Xu2021AnalysisApproach}
G.~Xu, G.~V. Eleftheriades, and S.~V. Hum, ``{Analysis and Design of General Printed Circuit Board Metagratings With an Equivalent Circuit Model Approach},'' \emph{IEEE Trans. Antennas Propag.}, vol.~69, no.~8, pp. 4657--4669, 2021.

\bibitem{Casolaro2019DynamicMetagratings}
A.~Casolaro, A.~Toscano, A.~Al{\`u}, and F.~Bilotti, ``Dynamic beam steering with reconfigurable metagratings,'' \emph{IEEE Trans. Antennas Propag.}, vol.~68, no.~3, pp. 1542--1552, 2019.

\bibitem{Popov2021NonManipulations}
V.~Popov, B.~Ratni, S.~N. Burokur, and F.~Boust, ``Non‐local reconfigurable sparse metasurface: Efficient near‐field and far‐field wavefront manipulations,'' \emph{Adv. Opt. Mater.}, vol.~9, p. 2001316, 2 2021.

\bibitem{Xu2022WideAntenna}
G.~Xu, G.~V. Eleftheriades, and S.~V. Hum, ``Wide-angle beam-steering and adaptive impedance matching with reconfigurable nonlocal leaky-wave antenna,'' \emph{IEEE Open J. Antennas Propag.}, vol.~3, pp. 1141--1153, 2022.

\bibitem{Biniashvili2021Eliminating}
L.~Biniashvili and A.~Epstein, ``Eliminating reflections in waveguide bends using a metagrating-inspired semianalytical methodology,'' \emph{IEEE Trans. Antennas Propag.}, vol.~70, no.~2, pp. 1221--1235, 2021.

\bibitem{Killamsetty2021MetagratingsExperiment}
V.~K. Killamsetty and A.~Epstein, ``Metagratings for perfect mode conversion in rectangular waveguides: Theory and experiment,'' \emph{Phys. Rev. Appl.}, vol.~16, no.~1, p. 014038, 2021.

\bibitem{Popov2020ConformalManipulation}
V.~Popov, S.~N. Burokur, and F.~Boust, ``Conformal sparse metasurfaces for wavefront manipulation,'' \emph{Phys. Rev. Appl.}, vol.~14, p.~1, 2020.

\bibitem{Xu2022ExtremeOptimization}
G.~Xu, V.~G. Ataloglou, S.~V. Hum, and G.~V. Eleftheriades, ``{Extreme Beam-Forming With Impedance Metasurfaces Featuring Embedded Sources and Auxiliary Surface Wave Optimization},'' \emph{IEEE Access}, pp. 28\,670--28\,684, 2022.

\bibitem{Kerzhner2022MetagratingApplications}
Y.~Kerzhner and A.~Epstein, ``Metagrating-assisted high-directivity sparse regular antenna arrays for scanning applications,'' \emph{IEEE Trans. Antennas Propag.}, vol.~71, pp. 650--659, 1 2023.

\bibitem{Hu2023LargeSwitching}
F.~Hu and A.~Epstein, ``Large-aperture cavity-excited metagrating antennas for dynamic beam switching,'' in \emph{2023 International Workshop on Antenna Technology (iWAT)}.\hskip 1em plus 0.5em minus 0.4em\relax IEEE, 2023, pp. 1--4.

\bibitem{Hu2023CavityPatterns}
------, ``Cavity-excited metagrating antennas with controlled frequency diverse patterns,'' in \emph{2023 IEEE International Symposium on Antennas and Propagation and USNC-URSI Radio Science Meeting (USNC-URSI)}.\hskip 1em plus 0.5em minus 0.4em\relax IEEE, 2023, pp. 1251--1252.

\bibitem{Yashno2022LargeAbsorption}
Y.~Yashno and A.~Epstein, ``Large-period multichannel metagratings for broad-angle absorption,'' in \emph{The 16th International Congress on Artificial Materials for Novel Wave Phenomena (Metamaterials)}.\hskip 1em plus 0.5em minus 0.4em\relax IEEE, 2022, pp. X--134.

\bibitem{Yashno2023BroadDiffusers}
------, ``Broad-angle multichannel metagrating diffusers,'' \emph{IEEE Trans. Antennas Propag.}, vol.~71, pp. 2409--2420, 3 2023.

\bibitem{Tan2023DesignExperiment}
Z.~Tan, J.~Yi, Q.~Cheng, and S.~N. Burokur, ``Design of perfect absorber based on metagratings: Theory and experiment,'' \emph{IEEE Trans. Antennas Propag.}, vol.~71, no.~2, pp. 1832--1842, 2023.

\bibitem{Ra'di2018reconfigurable}
Y.~Ra’di and A.~Al{\`u}, ``{Reconfigurable Metagratings},'' \emph{ACS Photon.}, vol.~5, no.~5, pp. 1779--1785, 2018.

\bibitem{Shklarsh2024SemianalyticallyMetagratings}
Y.~Shklarsh and A.~Epstein, ``Semianalytically designed dual-polarized printed-circuit-board (pcb)-compatible metagratings,'' \emph{IEEE Trans. Antennas Propag.}, vol.~72, no.~2, 2024.

\bibitem{Nguyen2024DesignReflection}
M.~A. Nguyen and G.~Byun, ``Design of anisotropic metagratings for diverse polarization anomalous reflection,'' in \emph{Proc. 18th Eur. Conf. Antennas Propag. (EuCAP)}, 2024.

\bibitem{Rahmanzadeh2020PerfectMetagrating}
M.~Rahmanzadeh and A.~Khavasi, ``{Perfect Anomalous Reflection Using a Compound Metallic Metagrating},'' \emph{Opt. Express}, vol.~28, no.~11, pp. 16\,439--16\,452, 2020.

\bibitem{Rajabalipanah2021AnalyticalMetagratings}
H.~Rajabalipanah and A.~Abdolali, ``{Analytical Design for Full-space Spatial Power Dividers using Metagratings},'' \emph{J. Opt. Soc. Am. B}, vol.~38, no.~10, pp. 2915--2919, 2021.

\bibitem{Rabinovich2020Dual-PolarizedReflection}
O.~Rabinovich and A.~Epstein, ``{Dual-polarized All-metallic Metagratings for Perfect Anomalous Reflection},'' \emph{Phys. Rev. Appl.}, vol.~14, no.~6, p. 064028, 2020.

\bibitem{Wang2021NovelMetasurfaces}
Y.~Wang and Y.~Ge, ``{Novel Single/dual Circularly Polarized Antennas Based on Polarization-conversion Reflective metasurfaces},'' \emph{Prog. Electromagn. Res. C}, vol. 108, pp. 237--25, 2021.

\bibitem{Guo2016AdvancesApplications}
C.~Guo, F.~Liu, S.~Chen, C.~Feng, and Z.~Zeng, ``Advances on exploiting polarization in wireless communications: Channels, technologies, and applications,'' \emph{IEEE Commun. Surv. Tutor.}, vol.~19, no.~1, pp. 125--166, 2016.

\bibitem{Ameri2019UltraMetasurfaces}
E.~Ameri, S.~H. Esmaeli, and S.~H. Sedighy, ``{Ultra Wideband Radar Cross Section Reduction by Using Polarization Conversion Metasurfaces},'' \emph{Sci. Rep.}, vol.~9, no.~1, p. 478, 2019.

\bibitem{Asadchy2016PerfectMetasurfaces}
V.~S. Asadchy, M.~Albooyeh, S.~N. Tcvetkova, A.~D{\'\i}az-Rubio, Y.~Ra'di, and S.~Tretyakov, ``{Perfect Control of Reflection and Refraction Using Spatially Dispersive Metasurfaces},'' \emph{Physical Review B}, vol.~94, no.~7, p. 075142, 2016.

\bibitem{Yepes2021PerfectSlab}
C.~Yepes, M.~Faenzi, S.~Maci, and E.~Martini, ``{Perfect Non-specular Reflection with Polarization Control by Using a Locally Passive Metasurface Sheet on a Grounded Dielectric Slab},'' \emph{Appl. Phys. Lett.}, vol. 118, no.~23, p. 231601, 2021.

\bibitem{Elad2023AnisotropicReflection}
S.~Elad and A.~Epstein, ``{Anisotropic Metagratings for Simultaneous Polarization Conversion and Anomalous Reflection},'' in \emph{Proc. 17th Eur. Conf. Antennas Propag. (EuCAP)}, 2023.

\bibitem{Note1}
While the model can be formulated for an arbitrary number of MAs as in \cite {Rabinovich2019ArbitraryMetagratings}, we focus here, as a fundamental case study, on anisotropic MGs with precisely two MAs per period. As shall be shown in Section \ref {subsec_PolConv_&_AnoRef}, these provide the minimal number of DOFs required for realizing the desired simultaneous polarization conversion and beam deflection.

\bibitem{Pozar2011MicrowaveEngineering}
D.~M. Pozar, \emph{Microwave Engineering}.\hskip 1em plus 0.5em minus 0.4em\relax John wiley \& sons, 2011.

\bibitem{Felsen1973RadiationWaves}
{Felsen, L. B. and Marcuvitz, N.}, \emph{{Radiation and Scattering of Waves}}.\hskip 1em plus 0.5em minus 0.4em\relax IEEE Press, 1973.

\bibitem{Note2}
\textcolor {black}{Note that due to the anisotropic nature of the rotated MAs, the incident field generally induces dipoles with non-vanishing projections on both the $x$ and $y$ axes, namely, generating scattered FB modes of both TE and TM polarizations. This is true even if the incident wave is strictly single-polarized (i.e., even if $E_\protect \mathrm {in}^\protect \mathrm {TE}=0$ or $E_\protect \mathrm {in}^\protect \mathrm {TM}=0$), as shall be demonstrated in Section \ref {sec_Results_and_Discussion}.}

\bibitem{Note3}
As in \cite {Rabinovich2019ArbitraryMetagratings}, appropriate DOF values for thus realizing a prescribed scattering functionality are found by running the MATLAB library function $\protect \mathtt {lsqnonlin}$ ca. 50 times with random initial values.

\bibitem{Awange2018OverdeterminedSystems}
J.~L. Awange, B.~Pal{\'a}ncz, R.~H. Lewis, and L.~V{\"o}lgyesi, ``{Overdetermined and Underdetermined Systems},'' \emph{Mathematical Geosciences: Hybrid Symbolic-Numeric Methods}, pp. 77--110, 2018.

\bibitem{Note4}
Herein, there is no distinction between regions $2^-$ and $2^+$ as $h_1\protect \!=\protect \!h_2$, hence the region between the PEC and the air-dielectric interface is simply denoted by $2$.

\bibitem{Balanis2016AntennaDesign}
C.~A. Balanis, \emph{Antenna Theory: Analysis and Design}.\hskip 1em plus 0.5em minus 0.4em\relax John wiley \& sons, 2016.

\bibitem{ROGERS}
J.~Coonrod, ``{Critical Aspects of Dielectric Constant Properties for High Frequency Circuit Design},'' \url{https://www.microwavejournal.com/ext/resources/Webinars/2015a/MWJ-Design-Dk-webinar-Dec-2015-rev3.pdf/}.

\bibitem{Note5}
\textcolor {black}{Note, that the simulated $\eta _c$ is the value of the relevant entry in the scattering matrix of the devised configuration, defined and simulated in CST. This scattering parameter, quantifying the scattered TM power coupled to the desired TM $m=-1$ FB mode relative to the overall incident TE power, is a part of the standard output by CST within such a Floquet port simulation.}

\bibitem{Note6}
\textcolor {black}{Indeed, the experimentally evaluated efficiency of $85.9\%$ for the anomalous reflection and polarization conversion MG devised herein is comparable with other measured beam manipulating MG efficiencies of $\sim 90\%$ reported in the past \cite {Rabinovich2019ExperimentalMetagratings}.}

\bibitem{Note7}
\textcolor {black}{Being only theoretical and neglecting loss, it is somewhat difficult to compare the solutions in \cite {Asadchy2016PerfectMetasurfaces, Yepes2021PerfectSlab} to the designs presented herein. If losses are neglected for the design in Section \ref {subsec_Perfect_TE-to-TM}, the efficiency reaches $99.4\%$, almost identical to the optimal unitary MS solutions. For the simulated designs herein and in \cite {Yepes2021PerfectSlab}, one may also comment on the $90\%$ fractional bandwidth (with respect to the peak efficiency), finding a bandwidth of about $5\%$ for the results presented in Fig. \ref {Fig_freqres9} of Section \ref {subsec_Experiment} compared to about $2\%$ in \cite {Yepes2021PerfectSlab}.}

\bibitem{NIST:DLMF}
``{\it NIST Digital Library of Mathematical Functions},'' \url{https://dlmf.nist.gov/}, Release 1.1.12 of 2023-12-15, f.~W.~J. Olver, A.~B. {Olde Daalhuis}, D.~W. Lozier, B.~I. Schneider, R.~F. Boisvert, C.~W. Clark, B.~R. Miller, B.~V. Saunders, H.~S. Cohl, and M.~A. McClain, eds.

\end{thebibliography}

\end{document}